\documentclass[amssymb,amsmath,superscriptaddress,reprint,twocolumn]{revtex4-1}
\usepackage{amsmath,diagbox,natbib,graphicx,subfigure,subeqnarray,fancyhdr,amsmath,multirow,color, mathrsfs,relsize}
\begin{document}

\hyphenation{form-ulated} \hyphenation{co-hesion}

\definecolor{greenseb}{RGB}{0,142,0}

\newcommand{\ek}[1]{\textcolor{red}{ek: #1}}
\newcommand{\sm}[1]{\textcolor{greenseb}{SM: #1}}

\newcommand{\red}[1]{\textcolor{red}{#1}}

\let\vaccent=\v 
\renewcommand{\v}[1]{\ensuremath{\mathbf{#1}}} 
\newcommand{\gv}[1]{\ensuremath{\mbox{\boldmath$ #1 $}}}  
\newcommand{\uv}[1]{\ensuremath{\mathbf{\hat{#1}}}} 
\newcommand{\abs}[1]{\left| #1 \right|} 
\newcommand{\avg}[1]{\left< #1 \right>} 
\newcommand{\grad}[1]{\gv{\nabla} #1} 
\let\divsymb=\div 
\renewcommand{\div}[1]{\gv{\nabla} \cdot #1} 
\newcommand{\curl}[1]{\gv{\nabla} \times #1} 
\newcommand{\eq}[1]{Equation~(\eqref{#1})}
\newcommand{\fig}[1]{Figure~\ref{#1}}
\newcommand{\pd}[2]{\frac{\partial #1}{\partial #2}}
\newcommand{\pdd}[2]{\frac{\partial^2 #1}{\partial #2^2}}
\newcommand{\diffd}{\mathrm{d}}
\newcommand{\dd}[2]{\frac{\diffd #1}{\diffd #2}}
\newcommand{\ddd}[2]{\frac{\diffd^2 #1}{\diffd #2^2}}
\newcommand{\boldnabla}{\mbox{\boldmath$\nabla$}}
\newcommand{\bolddelta}{\mbox{\boldmath$\delta$}}
\newcommand{\conj}[1]{\overline{#1}}
\newcommand{\real}[1]{\mathrm{Re} \left[ #1 \right]}
\newcommand{\imag}[1]{\mathrm{Im} \left[ #1 \right]}
\newcommand{\order}[1]{\mathcal O \left( #1 \right)}
\newcommand{\eee}[1]{\mathrm{e}^{ #1 }}
\newcommand{\ii}{\mathrm{i}} 

\renewcommand{\labelitemi}{$-$}

\newcommand{\ee}{\mathrm{e}}\newcommand{\ci}{\mathrm{i}}\newcommand{\ib}{\mathbf{i}}
\newcommand{\jb}{\mathbf{j}}\newcommand{\kb}{\mathbf{k}}\newcommand{\ab}{\mathbf{a}}
\newcommand{\Fb}{\mathbf{F}}\newcommand{\fb}{\mathbf{f}}\newcommand{\Gb}{\mathbf{G}}
\newcommand{\rhob}{\boldsymbol\rho}
\newcommand{\Mb}{\mathbf{M}Ä}\newcommand{\nb}{\mathbf{n}}\newcommand{\Sb}{\mathbf{S}}
\newcommand{\Sbs}{\mathbf{S^*}}\newcommand{\Rb}{\mathbf{R}}\newcommand{\Sigb}{\boldsymbol{\Sigma}}
\newcommand{\Sigbs}{\boldsymbol{\Sigma^*}}\newcommand{\alphab}{\boldsymbol\alpha}
\newcommand{\omegab}{\boldsymbol{\omega}}
\newcommand{\Omegab}{\boldsymbol{\Omega}}
\newcommand{\epsb}{\boldsymbol{\epsilon}}
\newcommand{\ub}{\mathbf{u}}
\newcommand{\eb}{\mathbf{e}}\newcommand{\vv}[1]{\underline{#1}}\newcommand{\ev}{\vv{e}}
\newcommand{\rv}{\vv{r}}\newcommand{\TT}[1]{\underline{\underline{#1}}}\newcommand{\omb}{\mathbf{\omega}}
\newcommand{\Ub}{\mathbf{U}}\newcommand{\xb}{\mathbf{x}}\newcommand{\pb}{\mathbf{p}}\newcommand{\rb}{\mathbf{r}}
\newcommand{\ssb}{\mathbf{s}}\newcommand{\Xb}{\mathbf{X}}\newcommand{\Pe}{\mbox{Pe}}
\newcommand{\mean}[1]{\langle #1\rangle}
\newcommand{\ddp}{[p]^\pm}\newcommand{\taub}{\mbox{\boldmath$\tau$}}\newcommand{\Fr}{\mbox{\textit{Fr}}}
\let\grad\nabla\newcommand{\z}{\zeta}\newcommand{\kk}{\kappa}\newcommand{\tkk}{\tilde{\kappa}}
\newcommand{\e}{\varepsilon}\newcommand{\zb}{\bar{\zeta}}\let\grad\nabla\let\bcdot\cdot
\newcommand{\half}{{\textstyle\frac{1}{2}}}
\newcommand{\textfrac}[2]{{\textstyle\frac{#1}{#2}}}
\newcommand{\LF}[1]{{#1}^{\mathrm{LF}}}\newcommand{\Lap}[1]{{#1}^{\mathrm{L}}}
\newcommand{\ds}{*\!*}\newcommand{\cond}[2]{\frac{\mathrm{D} #1}{\mathrm{D} #2}}
\newcommand{\pard}[2]{\frac{\partial #1}{\partial #2}}\newcommand{\totd}[2]{\frac{\mathrm{d}#1}{\mathrm{d}#2}}
\newcommand{\pardd}[3]{\frac{\partial^2 #1}{\partial #2 \partial #3}}
\newcommand{\Rey}{\mbox{Re}}\newcommand{\Imag}{\mbox{Im}}
\newcommand{\Fpint}{=\!\!\!\!\!\!\!\int}
\newcommand{\txi}{\tilde\xi}\newcommand{\dxi}{\delta\xi}
\newcommand{\tpsi}{\tilde\psi}\newcommand{\dpsi}{\delta\psi}
\makeatletter
\def\sgn{\mathop{\operator@font sgn}}
\DeclareRobustCommand{\threevdots}{%
  \vbox{
    \baselineskip2\p@\lineskiplimit\z@
    \kern-\p@
    \hbox{.}\hbox{.}\hbox{.}
  }}
\makeatother

\title{Phoretic \& hydrodynamic interactions of weakly-confined autophoretic particles} 
\author{Eva Kanso}
\email{kanso@usc.edu}
\affiliation{Aerospace and Mechanical Engineering, University of Southern California, Los Angeles, CA 90089-1191}
\author{S\'ebastien Michelin}
\email{sebastien.michelin@ladhyx.polytechnique.fr} 
\affiliation{LadHyX -- D\'epartement de M\'ecanique, Ecole Polytechnique -- CNRS, 91128 Palaiseau, France} 
\date{\today}
\begin{abstract}
Phoretic particles self-propel using self-generated physico-chemical gradients at their surface. Within a suspension, they interact hydrodynamically by setting the fluid around them into motion, and chemically by modifying the chemical background seen by their neighbours. While most phoretic systems evolve in confined environments due to buoyancy effects, most models focus on their interactions in unbounded flows. Here, we propose a first model for the interaction of phoretic particles in Hele-Shaw confinement and show that in this limit, hydrodynamic and phoretic interactions share not only the same scaling but also the same form, albeit in opposite directions. In essence, we show that phoretic interactions effectively reverse the sign of the interactions that would be obtained for swimmers interacting purely hydrodynamically. Yet, hydrodynamic interactions can not be neglected as they significantly impact the magnitude of the interactions.  This model is then used to analyse the behaviour of a suspension. The suspension exhibits swirling and clustering collective modes dictated by the orientational interactions between particles, similar to hydrodynamic swimmers, but here governed by the surface properties of the phoretic particle; the reversal in the sign of the interaction tends to slow down the swimming motion of the particles.
\end{abstract}

\maketitle


\section{Introduction}
To self-propel autonomously at the microscopic scale, biological and synthetic micro-swimmers must overcome the viscous resistance of the surrounding fluid to create asymmetric and non-reciprocal flow fields in their immediate vicinity~\citep{lauga2009}. Beyond their individual self-propulsion, understanding their interactions and collective behaviour is fascinating researchers across disciplines, particularly because their small scale suggest simpler interaction routes than larger and more complex organisms or systems.

While biological swimmers mainly rely on the actuation of flexible appendages such as flagella or cilia~\citep{brennen1977,lauga2016}, artificial microswimmers fall within two main categories. Externally-actuated swimmers respond to an external force or torque applied by a  magnetic~\citep{dreyfus2005}, electric~\citep{bricard2013} or acoustic field~\citep{wang2012} at the macroscopic level. In contrast, autophoretic (or fuel-based) swimmers exploit the chemical and electrical properties of their surface to generate slip flows within a thin interaction layer in response to local self-generated gradients of their physico-chemical environment. This ability to turn such gradients into fluid motion is known as \emph{phoresis}~\citep{anderson1989} and can arise from concentration of a diffusing solute species (diffusiophoresis), temperature (thermophoresis) or electric potential (electrophoresis). Popular experimental realizations of such systems include metallic or bi-metallic colloids catalyzing the decomposition of hydrogene peroxide solutions~\citep{paxton2004,howse2007,theurkauff2012,palacci2013,brown2014} or other redox reactions~\citep{ibele2009}, as well as heat-releasing particles in binary mixtures. It should be noted that in many aspects, active droplets, that achieve self-propulsion through self-generated Marangoni flows~\cite{izri2014,herminghaus2014}, can also be considered as examples of synthetic fuel-based swimmers. 

Because their individual self-propulsion is based solely on the interaction with their immediate microscopic environment, these systems have recently been intensely studied, experimentally and numerically, as canonical examples of active matter to analyse collective behaviour at the micron scale, demonstrating complex or chaotic behaviour as well as clustering~\citep{buttinoni2013,colberg2017,ilien2017,ginot2018,varma2018}. {Chemically-patterned systems (e.g. Janus particles) have played a central role in these investigations. Their chemical polarity establishes the chemical gradient required for propulsion, and can be obtained using either passive colloid partially-coated with a catalytic~\citep{howse2007,ke2010,valadares2010,ebbens2011,brown2014} or heat-absorbing layer~\citep{volpe2011,baraban2013}, bi-metallic swimmers~\citep{paxton2004,wang2006,theurkauff2012,davieswykes2017} or active material encapsulated in a passive colloid~\citep{palacci2013}. Beyond the qualitative understanding of the link between their asymmetry and their self-propulsion, the details of the competing physico-chemical mechanisms at the heart of each system are still the focus of ongoing investigations~\citep{moran2017,ibrahim2017}.}

All such systems share a common feature: they generate flow fields and motility from the dynamics of a physico-chemical fields (which will be taken in the following as the concentration of a chemical solute for simplicity and generality), that can diffuse and potentially be advected by the fluid flows. This provides two distinct interaction routes between individual swimmers. Like their biological counterparts, their self-propulsion sets the surrounding fluid into motion which influences the trajectory of their neighbours. But, just as bacteria and other microorganisms perform chemotaxis in response to different chemical nutrient or waste compounds, these systems can also exhibit a chemotactic behaviour and respond to the chemical signals left by the other particles~\cite{saha2014,tatuleacodrean2018}. Understanding the role of each interaction route~\cite{soto2014,liebchen2015,liebchen2017} as well as their interplay and competition~\cite{huang2017} is the focus of active research.

A major difficulty in the quantitative rationalization of experimental results on collective dynamics of such particles lies in their geometric environment. Most phoretic swimmers are either denser or lighter than the surrounding fluid and buoyancy forces effectively confine them to the immediate vicinity of a solid wall or a free-surface~\cite{palacci2010,theurkauff2012,palacci2013}. This confinement and reduced-dimensionality can have profound consequences on their collective behaviour~\cite{kruger2016}. Yet, the vast majority of existing models consider their evolution in a 3D unbounded environment~\cite{golestanian2007,julicher2009,michelin2014,moran2017}. Recent and successful attempts have demonstrated the ability of confining boundaries to  influence significantly the dynamics of single particles~\citep{uspal2014,uspal2016,mozaffari2016,malgaretti2018}, even providing controlling strategies for guidance~\citep{das2015,simmchen2016,davieswykes2017}. Because they can profoundly affect the flow field or chemical concentration generated by the particle, confining boundaries also significantly modify the interactions among them~\cite{dominguez2016,kruger2016}. As an example, their ability to screen differently the viscous and potential components of the flow field generated by the swimmer~\citep{blake1974} allows confining boundaries to significantly alter the clustering dynamics of such particles; these modifications  were recently shown to depend fundamentally on the number and nature of the confining surfaces~\citep{thutupalli2018}. 

The goal of the present work is to analyse the fundamental role of confinement on the interactions of many phoretic particles, and in particular on the relative weight of the hydrodynamic and chemical (phoretic) coupling. To this end, we model a dilute suspension of weakly-confined particles in a Hele-Shaw cell, effectively assuming a separation of three length scales:  the size of the swimmers, the depth of the confining chamber, and the typical distance between two swimmers. {Because of this separation of the three length scales, confinement profoundly modifies the interaction dynamics, which is driven by the hydrodynamic and chemical field screened by the presence of the confining walls, while the individual self-propulsion remains essentially unchanged at leading order. This effective decoupling of self-propulsion and interaction dynamics, which are normally intrinsically linked as they arise from the same slip distribution at the particles' surface, provides some fundamental insight on the influence of boundaries on the latter.}

After reviewing the fundamental mechanisms of self-propulsion of spherical Janus particles, the hydrodynamic and chemical signatures of individual particles as well as the resulting drifts in external flows and concentration gradients in Section~\ref{sec:unbounded}, we derive the leading order far-field hydrodynamic and chemical signatures of an individual particle in a Hele-Shaw environment in Section~\ref{sec:confined}. In both settings, the fundamental scalings of these signatures and interaction drifts are clearly identified and demonstrate the fundamental role of confinement in setting these interactions. Section~\ref{sec:2particles} presents the equations of motion for $N$ interacting particles and Section~\ref{sec:suspensions} finally applies this model to the dynamics of a dilute suspension. We finally discuss our results and present some perspectives in Section~\ref{sec:conclusions}.


\section{Janus particles in unbounded domains}
\label{sec:unbounded}

\begin{figure}[t]
\centerline{\includegraphics[scale=1]{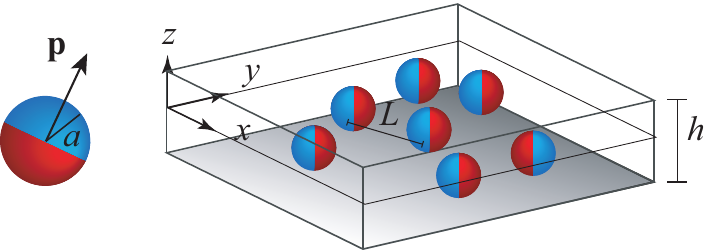}}
\caption{
\footnotesize  Schematics of a  single auto-phoretic Janus particle of radius $a$ in unconfined 3D space, with self-propelled velocity $U\mathbf{p}$, and a population of particles in weak Hele-Shaw confinement, where $a\ll h\ll L$. {The unit vector $\mathbf{p}$ indicating the orientation of the Janus particle is directed orthogonally to the plane that delineates the two sides of the Janus particle, depicted in red and blue respectively (color online). Specifically, $\mathbf{p}$ is oriented from red to blue.}
\label{fig:schematic}}
\end{figure}

We first analyse the individual motion, signature and interactions of chemically-active spherical particles in the absence of any confinement. The particles considered throughout the paper are spherical and chemically-active with a radius $a$. Their self-propulsion along self-generated gradients of a physico-chemical field results from their polar chemical activity and their sensitivity to the field~\citep{golestanian2007,julicher2009,moran2017}. In the following, we focus for simplicity on neutral diffusiophoresis for which the driving field is the concentration of a solute species produced or consumed at the surface of the particle. The principles and quantitative results can be easily adapted to other phoretic mechanisms such as electrophoresis or thermophoresis.

Before focusing on the interactions of such particles in confined environments, we first review their self-propulsion and interactions in the canonical context of unbounded flows. Most of the analysis presented in this section therefore summarizes some classical results on the self-propulsion~\cite{golestanian2007,julicher2009}, hydrodynamic coupling~\citep{kim2005} and phoretic interactions~\citep{saha2014,tatuleacodrean2018} of such particles in bulk flows,  and their derivation is briefly outlined here so as to emphasize their physical origin and relative scaling, which is critical to the later understanding of the screening effects in Section~\ref{sec:confined}.

The chemically-active spherical particles have radius $a$, and their physico-chemical properties are characterized by a chemical activity {$A(\xb_s)$ (i.e., their ability to modify the solute concentration) and mobility $M(\xb_s)$ (i.e., their ability to drive a surface slip flow from a local concentration gradient), that depend on the position $\xb_s$ along the surface of the particle}.  
Axisymmetric spherical particles are considered with an axis of symmetry $\pb$ (see figure~\ref{fig:schematic}).

\subsection{Self-propulsion of isolated auto-phoretic particles}

 The concentration field $C(\rb)$ around the particle relative to its background value satisfies the following system of equations:
\begin{align}
D&\nabla^2 C =0, \qquad \textrm{for   } r \geq a,\\
\left. C \right|_{r\rightarrow\infty} &=0,\qquad \left. D (\nb\cdot\grad C) \right|_{r = a} =-A(\mu). 
\end{align}
Here, $\rb$ is the position vector measured from the particle center, $r = \| \rb\|$ its magnitude, and $\mu=\cos\theta=\pb\cdot{\rb}/{r}$ in spherical polar coordinates given by $(r,\theta)$. This problem can be canonically solved for the concentration field~\citep{golestanian2007}:
\begin{equation}
\label{eq:C3d}
C(r,\mu)=\sum_{m=0}^\infty \frac{aA_m}{D(m+1)}\left(\frac{a}{r}\right)^{m+1}L_m(\mu),
\end{equation}
where $A_m=\frac{2m+1}{2}\int_{-1}^1A\,L_m\diffd \mu$ are the Legendre moments {of the axisymmetric activity distribution $A(\mu)$}, with $L_m(\mu)$ being the $m^{\rm th}$ Legendre polynomial.  
 
The gradient in chemical concentration induces a slip velocity at the surface of the phoretic particle given by~\citep{anderson1989}
\begin{equation}
 \ub_{\rm slip}=M(\mu)(\mathbf{I}-\nb\nb)\cdot \left. \grad C\right|_{r=a},
 \end{equation}
where $M(\mu)$ is the axisymmetric particle mobility, $\mathbf{I}$ is the identity tensor, and $\mathbf{n}\mathbf{n}$ refers to the dyadic product of the vector $\mathbf{n}$ with itself. This chemically-induced slip serves as a forcing boundary condition to solve for the Stokes flow around the force-and torque-free phoretic particle, 
\begin{equation}
\label{eq:stokes}
\eta \nabla^2 \mathbf{u} -\nabla p = 0,  \qquad \grad\cdot\mathbf{u} = 0, \qquad \textrm{for   } r \geq a,
\end{equation}
subject to boundary conditions  
\begin{equation}
 \left. \mathbf{u}\right|_{r \to \infty}   = 0, \qquad  \left. \mathbf{u}\right|_{r = a}  = \Ub + \ub_{\rm slip}.
\end{equation}
In Eq.~\eqref{eq:stokes}, $\ub$ is the fluid velocity field, $p$ the pressure field, $\eta$ the dynamic viscosity and $\Ub=U\pb$ the swimming velocity of the particle.
The relative flow velocity at the surface is purely tangential, therefore the flow field $\ub$ in the lab frame is generically given by~\cite{blake1971,michelin2011}
\begin{align}
\ub&=-\frac{1}{r^3}\pd{\psi}{\mu}\rb-\frac{1}{r(1-\mu^2)}\pd{\psi}{r}\left(\frac{\rb\rb}{r^2}-\mathbf{I}\right)\cdot\pb,
\label{eq:udef}
\end{align}
where $\psi$ is the stream function, 
\begin{align}
\psi&(r,\mu)=\frac{Ua^2(1-\mu^2)}{2\bar{r}}\nonumber\\
&+\sum_{n\geq 2}\frac{(2n+1)a^2\alpha_n}{2n(n+1)}(1-\mu^2)L_n'(\mu)\left[\bar{r}^{-n}-\bar{r}^{2-n}\right]\label{eq:psi}.
\end{align}
with $\bar{r}=r/a$. The prime in $L'_n$ denotes derivative with respect to $\mu$. Substituting Eq.~\eqref{eq:psi} into Eq.~\eqref{eq:udef}, one gets that the first term in $\psi$ corresponds to a source dipole ($u\sim (a/r)^3$ for $r\gg a$) and the term of order $n$ in the infinite sum includes a potential singularity (gradient of a source dipole) and a viscous singularity (gradient of a Stokeslet). For example, $n=2$ includes a source quadrupole and force dipole, $n=3$ a source octopole and a force quadrupole, and so on. The intensity of these singularities is defined in terms of $\alpha_n$, given for $n\geq 2$ by
\begin{equation}
\alpha_n=-\frac{1}{2D}\sum_{m,p=0}^\infty\frac{A_mM_p}{m+1}\int_{-1}^1(1-\mu^2)L_m'L_n'L_p\diffd\mu,  
\end{equation}
where $M_p=\frac{2p+1}{2}\int_{-1}^1M\,L_p\diffd \mu$ are the Legendre moments of $M(\mu)$. 
The magnitude $U$ of the swimming velocity is given by~\citep{golestanian2007}:

\begin{equation}
U=-\sum_{m=1}^{\infty}\frac{mA_m}{2m+1}\left(\frac{M_{m-1}}{2m-1}-\frac{M_{m+1}}{2m+3}\right).
\end{equation}

The physico-chemical properties of the axisymmetric particle are set by its activity and mobility distribution $A(\mu)$ and $M(\mu)$. For a generic particle, these two functions are arbitrary (they can be positive or negative). A specific example is that of a hemispherical Janus particle, as shown in figure~\ref{fig:schematic}, with chemical properties of the front  half  given by $(A_f, M_f)$ and of the back half given by $(A_b,M_b)$. The first two Legendre moments of the activity and mobility distributions are given by $A_0=(A_f+A_b)/2$, $A_1=3(A_f-A_b)/4$, and $M_0=(M_f+M_b)/2$, $M_1=3(M_f-M_b)/4$, whereas all non-zero even moments are identically zero, that is to say,  $A_{2n}= M_{2n} = 0$ for $n\neq 0$. The swimming velocity $U$ and $\alpha_2$ can be found analytically~\cite{golestanian2007,michelin2014},
\begin{equation}
U=  -\frac{A_1M_0}{3D}= \frac{(A_b-A_f)(M_f+M_b)}{8D}, \label{eq:U3d}
\end{equation}
and
\begin{equation}
\alpha_2= -\dfrac{16\kappa A_1M_1}{9D} = -\frac{\kappa(M_f-M_b)(A_f-A_b)}{D} \label{eq:alpha2},
\end{equation}
with $\kappa$ a numerical constant, defined as
\begin{equation}
\kappa=\frac{3}{4}\sum_{m=1}^\infty\frac{2m+1}{m+1}\left[\int_0^1 L_m\mathrm{d}\mu\right]\left[\int_0^1 \mu(1-\mu^2)L_m'\mathrm{d}\mu\right],
\end{equation} 
such that $\kappa\approx 0.0872$.

\begin{table*}[!t]
  \begin{center}
  \begin{tabular}{l|ccc|cc}
   & \multicolumn{3}{c|}{\textbf{Hydrodynamic signature}} & \multicolumn{2}{c}{\textbf{Chemical signature}}\\[0.5ex] 
   \hline 
       &&&&& \\[-1.5ex]
    & \multicolumn{3}{c|}{ $\mathbf{u}(\mathbf{r})$ } & \multicolumn{2}{c}{$C(\mathbf{r})$}\\[0.5ex]  
  & force dipole & source dipole & force quadrupole  & source & source dipole \\[0.5ex] \hline 
  &&&&& \\[-0.75ex]
  intensity & $a^2 \dfrac{A_1M_1}{D}$& $a^3 \dfrac{A_1M_0}{D}$& $a^3 \alpha_3$& $a^2 \dfrac{A_0}{D}$ & $a^3 \dfrac{A_1}{D}$  \\[2.5ex]
  decay rate &$\dfrac{1}{r^2}$  & $\dfrac{1}{r^3}$ & $\dfrac{1}{r^3}$ & $\dfrac{1}{r}$ & $\dfrac{1}{r^2}$  
        \end{tabular}  
  \caption{Scaling laws of the  chemical and hydrodynamic fields  created by an isolated self-propelled phoretic particle in unbounded 3D domain based on Eqs.~\eqref{eq:chemicalfield_3d} and~\eqref{eq:velfield3d}, respectively. Here, $A_{0}, A_1$ and $M_{0},M_1$ represent the surface activity and motility of the particle,  $D$ is the molecular diffusivity of the solute, $a$ is the particle size, and $r$ is the distance from the particle at which these fields are evaluated. }
  \label{tab:signature3d}
  \end{center}
\end{table*}

\subsection{Hydrodynamic and chemical signatures}

The chemical and hydrodynamic signatures of a Janus particle in an unbounded three-dimensional (3D) domain are obtained by keeping the two most dominant terms in Eqs.~\eqref{eq:C3d} and~\eqref{eq:udef}, respectively. The \emph{chemical signature}  consists of (i) a source of solute proportional to the net production rate $A_0$ and (ii) a source dipole proportional to $A_1$, 
\begin{equation}
\label{eq:chemicalfield_3d}
C =  \dfrac{1}{4\pi} \left(\frac{4\pi a^2A_0}{D}\right)\dfrac{1}{r}+\dfrac{1}{4\pi}\left(\frac{2 \pi a^3A_1}{D}\right)\left(\frac{\pb\cdot\rb}{r^3}\right)+O\left(\dfrac{a^3}{r^3}\right).
\end{equation}

The \emph{Hydrodynamic signature} consists of (i) a force dipole (or stresslet) proportional to $\alpha_2$, (ii) a source dipole proportional to $U$ and (iii) a force quadrupole proportional to $\alpha_3$, 
\begin{equation}
\begin{split}
\ub=&\frac{1}{8\pi\eta}\grad\left(\frac{\mathbf{I}}{r}+\frac{\rb\rb}{r^3}\right):\mathbf{A} -\dfrac{1}{4\pi}\grad\left(\frac{\rb}{r^3}\right)\cdot  \mathbf{B} \\
&+\frac{1}{8\pi\eta}\grad\grad\left(\frac{\mathbf{I}}{r}+\frac{\rb\rb}{r^3}\right)\threevdots \, \mathbf{C} +O\left(\frac{a^4}{r^4}\right),\label{eq:velfield3d}
\end{split}
\end{equation}
where the coefficients are given by
\begin{align}
\mathbf{A} = \frac{10}{3}\pi\eta a^2\alpha_2(3\pb\pb-\mathbf{I}), \quad \mathbf{B} = 2\pi a^3U\pb ,  \label{eq:singularities1} \\
\mathbf{C} = \frac{7}{24}\pi\eta a^3\alpha_3\left[\pb\mathbf{I}+(\mathbf{I}\pb)^{T_{12}}+\mathbf{I}\pb-18\pb\pb\pb\right].\label{eq:singularities2}
\end{align}
Here, $(\cdot)^{T_{12}}$ is the transpose over the first two indices  (i.e. $(\mathbf{C})^{T_{12}}_{ijk}=C_{jik}$). 

Note that we consistently carry the development of $C$ and $\ub$ up to $O(a^3/r^3)$ and $O(a^4/r^4)$, respectively, since the chemical and hydrodynamic drift on neighboring particles will be driven by $\grad C$ and $\ub$, respectively. The leading-order terms in the fluid velocity field $\mathbf{u}(\mathbf{r})$ and chemical concentration field $C(\mathbf{r})$ are summarized in table~\ref{tab:signature3d}. We highlight that the axisymmetric Janus particle swims in the $\mathbf{p}$-direction at speed $U$ given in Eq.~\eqref{eq:U3d}  but it does not reorient under the influence of its own chemical activity. The results presented in Table~\ref{tab:signature3d} emphasize that the dominant hydrodynamic (force dipole) and chemical interactions (source) result in particles drift following the same algebraic decay in $1/r^2$ in the far-field limit. This underscores the necessity to account for both types of interactions, in contrast with simplifying assumptions regularly made in existing studies where hydrodynamic interactions are neglected~\cite{soto2015,liebchen2018b}, mostly on the grounds that the force dipole vanishes for hemispheric Janus particles, which is only correct for the very specific case of a uniform mobility~\citep{michelin2014}, which is likely difficult to achieve experimentally.

\subsection{Hydrodynamic and phoretic drift in external flows and chemical gradients}

When the particle is placed in an external and possibly non-uniform flow $\ub_\infty(\rb)$, it will translate and reorient.  For a spherical particle,  the translation and rotational drifts are given at the leading order by Faxen's laws~\citep{kim2005}:
\begin{equation}
\label{eq:drift}
\Ub_d=\ub_\infty(\rb_0),\qquad \Omegab_d=\frac{1}{2}\omega_\infty(\rb_0),
\end{equation}
where $\omega_\infty=\nabla\times\ub_\infty$ is the external vorticity field and $\rb_0$ is the position of the particle's centroid. Non-spherical particles are also sensitive to the local rate of strain due to their non-isotropy and thin elongated particles would reorient in the principal strain direction~\citep{Jeffery1922}.

The existence of an external non-uniform concentration field $C_\infty(\rb)$, generated for example by the particle's neighbors, induces an additional slip velocity on the particle's boundary that leads to a phoretic drift and reorientation~\citep{anderson1989,pohl2014}. To compute the phoretic drift velocities, one needs to solve the following solute diffusion problem
\begin{align}
\label{eq:Cdrift}
D&\nabla^2 C_d=0, \qquad \textrm{for   } r\geq a,\\[2ex]
\left. C_d\right|_{r\rightarrow \infty} & =C_\infty(\rb) \approx C_\infty^0+\Gb_\infty\cdot\rb+\frac{1}{2}\mathbf{H}_\infty:\rb\rb+\ldots,\\
& \left. D(\nb\cdot\grad C_d)\right|_{r = a} =0.
\end{align}
Here, we expanded the chemical field $C_\infty$ in a Taylor series about the particle position $\mathbf{r}_0$, with $\Gb_\infty = \left.  \grad C_\infty\right|_{\rb_0}$ and $\mathbf{H}_\infty = \left. \grad \grad C_\infty \right|_{\rb_0}$ evaluated at the particle location. The solution to Eq.~\eqref{eq:Cdrift} is uniquely obtained as~\citep{jackson1962} 
\begin{align}
C_d(\rb)=C_\infty^0&+(\Gb_\infty\cdot\rb)\left[1+\frac{1}{2}\left(\frac{a}{r}\right)^3\right]\nonumber\\
&+\frac{1}{2}(\mathbf{H}_\infty:\rb\rb)\left[1+\frac{2}{3}\left(\frac{a}{r}\right)^5\right] + \ldots
\end{align}
The additional slip velocity on the boundary of the particle is given by
\begin{align}
\ub_{\rm slip}&= M(\rb)(\mathbf{I}-\nb\nb)\cdot\grad C_d \nonumber\\
&=M(\rb)(\mathbf{I}-\nb\nb)\cdot\left[\frac{3}{2}\Gb_\infty+\frac{5}{3}\mathbf{H}_\infty\cdot\nb\right]+ \ldots
\end{align}
When the background concentration field $C_\infty$ is generated by other particles located at a distance $L$ far enough from the particle of interest, the
contribution of the second term involving $\mathbf{H}_\infty = \left. \grad\grad C_\infty\right|_{\rb_0}$ is subdominant by a factor $a/L$; we thus omit it in the following analysis. 

Determining the translation and rotation velocities of a force- and torque-free particle for a given prescribed slip distribution is a now-classical linear fluid dynamics problem~\citep{stone1996}. For a spherical particle, the reciprocal theorem for Stokes' flow around an isolated sphere provides the drift translational and rotational velocities as~\citep{anderson1989}

\begin{align}
\Ub_d^c&=-\left\langle\ub_\textrm{slip}(\rb)\right\rangle=-\frac{3\Gb_\infty}{2}\cdot\left\langle M(\rb)(\mathbf{I}-\nb\nb)\right\rangle,\\
\Omegab_d^c&=-\frac{3}{2a}{\left\langle\nb\times\ub_\textrm{slip}(\rb)\right\rangle}=-\frac{9}{4a}\left\langle M(\rb)\nb\right\rangle\times\Gb_\infty .
\end{align}
{where $\langle\hdots\rangle$ denote the spatial average taken over the particle's surface.} These expressions can be simplified further for axisymmetric particles noting that~\citep{tatuleacodrean2018}
\begin{align}
\langle M(\rb)\nb\nb\rangle
&=\frac{M_0}{3}\mathbf{I}+\frac{M_2}{15}(3\pb\pb-\mathbf{I}),
\label{eq:simplify1}\\
\langle M(\rb)\nb\rangle
&=\frac{M_1\pb}{3}.
\label{eq:simplify2}
\end{align}
so that the translational and rotational drift velocities can finally be rewritten as (with $\Gb_\infty=\left.\grad C_\infty\right|_{\rb_0}$), 
\begin{align}
\Ub_d^c&=-M_0\left.\grad C_\infty\right|_{\rb_0}+\frac{M_2}{10}(3\pb\pb-\mathbf{I})\cdot\left.\grad C_\infty\right|_{\rb_0}, \label{eq:trans_drift3d}\\ 
\Omegab_d^c&=-\frac{3M_1}{4a}\pb\times\left.\grad C_\infty\right|_{\rb_0}.\label{eq:rot_drift3d}
\end{align}

For the particular case of a hemispheric Janus particle, all even modes $A_{2n}$ and $M_{2n}$ (for $n>0$) are zero due to symmetry, and one finally obtains
\begin{equation}
\label{eq:chemdriftvel}
\Ub_d^c =- M_0\left.\grad C_\infty\right|_{\rb_0}, \qquad 
\Omegab_d^c =-\frac{3M_1}{4a}\,\pb\times\left.\grad C_\infty\right|_{\rb_0}.
\end{equation}

The phoretic interactions include a reorientation to align the particle with (resp. against) the chemical gradient if $M_1<0$ (resp. $M_1>0$), Eq.~\eqref{eq:rot_drift3d}, as well as a translational drift with components both along the chemical gradient and in the particle's direction, Eq.~\eqref{eq:trans_drift3d}. These different contributions provide different modes of chemotaxis to the catalytic colloids, that were recently analyzed in detail~\cite{saha2014,tatuleacodrean2018}.
The translational velocity scales as $U_{d}^c \sim M_0 G$ and the reorientation velocity scales as $\Omega_{d}^c \sim M_1 G/a$, where  $G$ the characteristic solute concentration gradient created by the other particles at the location of the particle considered.
The reorientation time scale  is $\tau_{\rm rotation}^c= 1/ \Omega_d^c \sim a/M_1 G$. \\
 
We now briefly discuss the relative magnitude of these chemical and hydrodynamic effects. 
When the background chemical gradient experienced by a given particle is due to a second particle located at a distance $L$, this gradient scales as $G\sim (A_0/D)(a/L)^2$ at leading order, if these particles are net sources or sinks of solute ($A_0\neq 0$), see Eq.~\eqref{eq:chemicalfield_3d}. Consequently, the chemical drift velocity scales as $U_d^c \sim (M_0 A_0/D)(a/L)^2$ and the reorientation time scale is $\tau_{\rm rotation}^c = 1/\Omega_d^c \sim a(D/A_0M_1)( L^2/a^2)$. If the net production rate vanishes ($A_0=0$), the gradients are weaker, scaling as $G\sim (A_1/D)(a/L)^3$, and therefore
$U_d^c \sim (M_0 A_1/D)(a/L)^3$ and $\tau_{\rm rotation}^c \sim a(D/A_1M_1)( L^3/a^3)$.

\begin{table*}[!t]
  \begin{center}    
         \begin{tabular}{l|ccc|cc}
   & \multicolumn{3}{c|}{\textbf{Hydrodynamic drift}} & \multicolumn{2}{c}{\textbf{Chemical drift}}\\[0.5ex] 
  & force dipole  & source dipole & force quadrupole & source & source dipole \\[0.5ex] \hline 
  &&&&& \\[-0.75ex]
translational drift $U_d$ & $\dfrac{a^2}{L^2} \dfrac{A_1M_1}{D}$ 
& $\dfrac{a^3}{L^3} \dfrac{A_1M_0}{D}$ & $\dfrac{a^3\alpha_3}{L^3}$
& $\dfrac{a^2}{L^2} \dfrac{A_0M_0}{D}$ & $\dfrac{a^3}{L^3} \dfrac{A_1M_0}{D}$  \\[2.5ex]
rotational drift $\Omega_d$ &$\dfrac{1}{L}\dfrac{a^2}{L^2} \dfrac{A_1M_1}{D}$    
& $\dfrac{1}{L}\dfrac{a^3}{L^3} \dfrac{A_1M_0}{D}$ & $\dfrac{1}{L}\dfrac{a^3\alpha_3}{L^3}$
& $\dfrac{1}{a}\dfrac{a^2}{L^2} \dfrac{A_0M_1}{D}$ & $\dfrac{1}{a}\dfrac{a^3}{L^3} \dfrac{A_1M_1}{D}$ \\[0.5ex] 
        \end{tabular}
          \caption{Scaling laws of the chemical and hydrodynamic drift in unbounded 3D domain created by a phoretic particle at a distance $L$ from the particle of interest.}
  \label{tab:drift3d}
  \end{center}
\end{table*}

These scalings for chemical interactions between particles can be compared to their hydrodynamic counterparts. The leading order translational hydrodynamic drift scales as $U_d^h \sim (M_1A_1/D) (a/L)^2$, see Eq.~\eqref{eq:velfield3d}, and the reorientation time scales as $\tau^h_{\rm rotation} \sim L (D/A_1M_1/)(L/a)^2$.  A summary of the scaling of the translational and rotational velocities due to  hydrodynamic and phoretic drifts is given in table~\ref{tab:drift3d}.

When the particles act as net sources or sinks of solute ($A_0\neq 0$), the chemical and hydrodynamic drift velocities are of the same order $U_d^c \sim U_d^h$ but chemical rotations act much faster than hydrodynamic rotations $\tau_{\rm rotation}^c \ll \tau_{\rm rotation}^h$.  
When the chemical signature of the particle is a source dipole only ($A_0=0$), the hydrodynamic drift is dominant  $U_d^c \ll U_d^h$ whereas chemical and hydrodynamic rotations are of the same order $\tau_{\rm rotation}^c \sim \tau_{\rm rotation}^h$.

\section{Weakly-confined Janus particles in Hele-Shaw cells}
\label{sec:confined}

Section~\ref{sec:unbounded} established that in an unbounded domain (i) the dynamics of individual Janus particles is the superposition of their self-propulsion and the translational and rotational drifts associated with the background concentration of solute (the latter is not related to their activity), (ii) only the latter plays a role in their re-orientation when the particles are net sources or sinks of solute. 

We investigate here how these results and associated scalings are modified when the Janus particles are confined in a Hele-Shaw cell consisting of two no-slip walls that are chemically inert, with no activity or mobility, and separated by a distance $h$ (Figure~\ref{fig:schematic}b). We consider the joint limit of (i) \emph{weak confinement}, where the particle radius $a$ is much smaller than the gap $h$ between the walls, namely, $a\ll h$, and (ii) \emph{Hele-Shaw interactions}, where the typical distance $L$ between particles  is much greater than the cell depth leading to dilute suspensions with $h\ll L$. The first condition implies that the chemical and hydrodynamic drift of a particle in response to the local chemical and hydrodynamic fields are obtained in the same way as those of a particle in unbounded flow. The second condition is important to determine these local fields given a dilute suspension of particles in the Hele-Shaw cell, as discussed next.


\subsection{Chemical signature in Hele-Shaw confinement}
\label{sec:chem_confined}

Within the Hele-Shaw cell and outside of the individual particles, the solute concentration is governed by Laplace's equation with no-flux condition at the confining walls:
\begin{equation}
\label{eq:heleshaw}
D\nabla^2 C=0, \qquad  \textrm{    with   } \left. \pd{C}{z}\right|_{z=\pm h/2}=0.
\end{equation}
Here, we consider the Cartesian coordinates $(x,y,z)$ such that the $(x,y)$-plane is located in the middle of the Hele-Shaw cell, parallel to the bounding walls, and $z$ is along its depth, and we introduce the corresponding  inertial frame $(\mathbf{e}_x,\mathbf{e}_y,\mathbf{e}_z)$ as depicted in figure~\ref{fig:schematic}. We non-dimensionalize length in the $z$-direction by $h$ and in the $(x,y)$-plane by $L$ with $\epsilon=h/L\ll 1$, and we seek a solution to Eq.~\eqref{eq:heleshaw} in the form $C=C_0+\epsilon^2 C_1+\ldots$. We immediately obtain from Eq.~\eqref{eq:heleshaw} that $C_0$ is independent of $z$. That is to say, at the leading order, the solute concentration is homogeneous across the channel depth and its leading order gradient is horizontal. A direct but fundamental consequence of this homogeneity is that all the Janus particles will be forced to align parallel to the $(x,y)$-plane, which justifies a quasi-2D approach to the present problem. 
The presence of a net source of solute in 2D confinement would induce a logarithmic far-field singular behavior in
the concentration field. For a well-posed problem, one needs either to consider two different populations of particles (net producers and net consumers in similar proportions) or a single population of dipolar particles. We focus on the latter.

We consider the solute concentration field generated by a single Janus particle located at $\mathbf{r}_0$ and oriented  horizontally such that $\pb\cdot\eb_z=0$. In an unbounded flow domain, the concentration generated by this chemical dipole is obtained by setting $A_0 = 0$ in Eq.~\eqref{eq:chemicalfield_3d}, which we rewrite as 
\begin{equation}
C=\frac{1}{4\pi}\left(\frac{2\pi a^3 A_1}{D}\right)\frac{\pb\cdot(\rb-\rb_0)}{\|\rb-\rb_0\|^3},
\end{equation} 
where  $2\pi a^3 A_1/D$ is the intensity of the chemical dipole in the unbounded domain. 
When the particle is located within a Hele-Shaw channel, at $\rb_0=\delta\eb_z$, where $\delta$ is the vertical position of the particle relative to the central plane ($|\delta|<h/2$), the concentration within the channel gap can be reconstructed using the method of images by superimposing identical dipoles located at vertical positions $z_{n}^+=\delta+2nh$ and $z_{n}^-=-\delta+(2n+1)h$ ($n\in \mathbb{Z}$). To this end, we write $\rb=\xb+z\eb_z$, where $\xb = x\mathbf{e}_x + y \mathbf{e}_y$. The concentration field due to the chemical dipole and its infinite image system is given by
\begin{align}
C=\frac{1}{4\pi}&\left(\dfrac{2\pi a^3 A_1}{D}\right)(\pb\cdot\xb)\nonumber\\
&\times\sum_{n=-\infty}^\infty\left\{\dfrac{1}{\|\rb-z^+_{n}\eb_z\|^{3/2}}+\dfrac{1}{\|\rb-z^-_{n}\eb_z\|^{3/2}}\right\}\cdot
\end{align}
In the Hele-Shaw limit $\|\xb\| \gg h$, the leading order concentration field is obtained using Riemann's sum as
\begin{align}
\label{eq:chemical2d}
C^{\rm H\textrm{-}S}  &= \ \dfrac{1}{4\pi}\left(\dfrac{2\pi a^3 A_1}{D}\right) \dfrac{(\pb\cdot\xb)}{h\|\xb\|^2}\sum_{m=-\infty}^\infty\frac{h/\| \xb \|}{\left[1+\left(\dfrac{mh}{\|\xb\|}\right)^2\right]^{3/2}} \nonumber\\
&= \ \dfrac{1}{2\pi} \left(\dfrac{2 \pi a^3 A_1}{Dh}\right)\dfrac{(\pb\cdot\xb)}{\|\xb\|^2}\cdot
\end{align}
This concentration field corresponds to a \emph{two-dimensional chemical dipole} of intensity $2\pi a^3 A_1/hD$. It is homogeneous within the channel depth, i.e., it is independent of the particle's vertical position $\delta$.

 
\begin{table*}[t]
  \begin{center}
  \begin{tabular}{l|ccc|cc}
   & \multicolumn{3}{c|}{\textbf{Hydrodynamic signature}} & \multicolumn{2}{c}{\textbf{Chemical signature}}\\[0.5ex] 
   \hline 
       &&& &&\\[-1.5ex]
    & \multicolumn{3}{c|}{ $\mathbf{u}^{\rm H\textrm{-}S}(\xb)$ } & \multicolumn{2}{c}{$C^{\rm H\textrm{-}S}(\xb)$}\\[0.5ex] 
  & force dipole & source dipole    & force quadrupole  & source & source dipole  \\[0.5ex] \hline 
  &&& &&\\[-0.75ex]
  intensity & $h a^2 \dfrac{A_1M_1}{D}$ & $\dfrac{a^3}{h} \dfrac{A_1M_0}{D}$ & $\dfrac{a^3\alpha_3}{h}$ & $-$ & $\dfrac{a^3}{h} \dfrac{A_1}{D}$   \\[2.5ex]
  decay rate &$\dfrac{1}{L^3}$  & $\dfrac{1}{L^2}$ & $\dfrac{1}{L^4}$ & $-$ & $\dfrac{1}{L}$  
        \end{tabular}
  \caption{Scaling laws of the chemical and hydrodynamic signatures  induced by a confined self-propelled phoretic particle in a Hele-Shaw cell of width $h$ such that $a\ll h$ and $h\ll L$. Parameter values are defined in table~\ref{tab:signature3d}. }
  \label{tab:signature2d}
  \end{center}
\end{table*}


\subsection{Hydrodynamic signature in Hele-Shaw confinement}
\label{sec:hydro_confined}

The far-field hydrodynamic signature of a Janus particle in an unbounded fluid domain is given by Eq.~\eqref{eq:velfield3d}. The leading order term of the velocity field is that of a force dipole (or Stokeslet dipole) whose velocity field decays as $1/r^2$. The dominant correction to the leading order term includes a potential horizontal source dipole and force quadrupole, both decaying as $1/r^3$. 

We now consider a Janus particle that is confined between the two no-slip surfaces. We are interested in its hydrodynamic signature in the (far-field) Hele-Shaw limit, at distances $L$ much greater than the channel depth $h$.   By linearity of the Stokes equations, the velocity field produced by the confined particle is the sum of the velocity fields produced by the confined singularities:  force dipole, potential source dipole and force quadrupole. 
The effect of confinement between two rigid walls on a force singularity (Stokeslet) is analyzed at length by  Liron and Mochon~\citep{liron1976}.  They showed that in the far-field ($L \gg h$),
 a Stokeslet oriented along the horizontal direction (parallel to the confining walls)  induces an exponentially-decaying velocity field in the $z$-direction, whereas in the $(x,y)$-plane its dominant behavior corresponds to a two-dimensional source dipole. 
The direction of the source dipole is in the same direction as the original Stokeslet and its strength depends in a parabolic way on its placement between the two walls.
Mathematically, for a Stokeslet of unit strength located in a plane $z = \delta$ between the two walls such that
\begin{equation}
\mathbf{u}_{\rm st} =\frac{1}{8\pi\eta} (\frac{\mathbf{I}}{r}+\frac{\rb\rb}{r^3})\cdot \mathbf{p},
\end{equation}
the leading order far-field flow in Hele-Shaw confinement is given by (see Eq.~(51) in Ref.~\cite{liron1976})
\begin{equation}
\label{eq:heleshaw_stokeslet}
\mathbf{u}_{\rm st}^{ \rm H\textrm{-}S} = - \dfrac{3h}{2\pi\eta}\left(\dfrac{1}{4}-\dfrac{\delta^2}{h^2}\right) \left(\dfrac{1}{4}-\dfrac{z^2}{h^2}\right) \left(\dfrac{\mathbf{I}}{\|\xb\|^2} - 
 2\dfrac{\xb\xb}{\|\xb\|^4} \right) \cdot\mathbf{p}.
\end{equation}
Therefore, the far-field flow generated by a horizontal force dipole corresponding to the first term in Eq.~\eqref{eq:velfield3d},
\begin{equation} 
\mathbf{u}_{\rm fd} = \frac{1}{8\pi\eta}\grad \left(\frac{\mathbf{I}}{r}+\frac{\rb\rb}{r^3}\right): \mathbf{A},
\end{equation}
becomes, when confined between two walls, that of a two-dimensional source quadrupole decaying as $1/\|\xb\|^3$,
\begin{align}
\mathbf{u}_{\rm fd}^{\rm H\textrm{-}S} = -\dfrac{3h}{2\pi\eta}&\left(\dfrac{1}{4}-\dfrac{\delta^2}{h^2}\right)\left(\dfrac{1}{4}-\dfrac{z^2}{h^2}\right)   \nonumber\\
&\times\grad \left(\dfrac{\mathbf{I}}{\|\xb\|^2} -  2\dfrac{\xb\xb}{\|\xb\|^4} \right):\mathbf{A}.
\label{eq:flowforcedipole2d_1}
\end{align}
where $\mathbf{A} =  10\pi\eta a^2\alpha_2(\pb\pb-\mathbf{I}/3)$, see Eq.~\eqref{eq:singularities1}, and $\alpha_2$  is given in Eq.~\eqref{eq:alpha2}.
Substituting into Eq.~\eqref{eq:flowforcedipole2d_1} and evaluating the resulting expression at $\delta = z = 0$, the leading order flow field generated by a horizontal force dipole is obtained as
\begin{equation}
\mathbf{u}_{\rm fd}^{\rm H\textrm{-}S} = \dfrac{5ha^2 \kappa A_1 M_1}{9D}  \left[ \grad \left(\dfrac{\mathbf{I}}{\|\xb\|^2} -  2\dfrac{\xb\xb}{\|\xb\|^4} \right)\right]:(3\pb\pb-\mathbf{I}).
\label{eq:flowforcedipole2d}
\end{equation}

We now examine the effect of confinement on the flow generated by a 3D source dipole. 
We recall that the source dipole can be seen either as a potential flow solution to the Stokes equations, associated with Laplace's equation~\citep{Spagnolie2012}, or as a degenerate force quadrupole. Following the latter approach, the source dipole contribution in the unconfined domain corresponding to the second term in Eq.~\eqref{eq:velfield3d} can be rewritten as the Laplacian of a Stokeslet,
\begin{equation}
\mathbf{u}_{\rm sd} =  - \dfrac{1}{4\pi} \nabla \left(\dfrac{\rb}{r^3}\right)\cdot \mathbf{B} 
=-\dfrac{1}{8\pi} \grad^2\left( \frac{\mathbf{I}}{r}+\frac{\rb\rb}{r^3}\right)\cdot \mathbf{B},
\end{equation}
where $\mathbf{B} = 2\pi a^3 U\mathbf{p}$ is given by Eq.~\eqref{eq:singularities1} and the last equality follows directly by differentiation [e.g. Kim \& Karrila~\citep{kim2005}, chapter 2, Eq.~(2.12)]. 
{The leading order velocity field associated with the confined potential singularity is then obtained by considering the dominant contribution to  the Laplacian of Eq.~\eqref{eq:heleshaw_stokeslet}, and} is given by 
\begin{align}
\mathbf{u}_{\rm sd}^{\rm H\textrm{-}S}  
&=-\dfrac{3}{\pi h} \left(\dfrac{1}{4}-\dfrac{\delta^2}{h^2}\right) \left(\dfrac{\mathbf{I}}{\|\xb\|^2} -  2\dfrac{\xb\xb}{\|\xb\|^4} \right) \cdot \mathbf{B}.
\end{align}
Evaluating at $\delta = 0$ and substituting the expression for $U$ from Eq.~\eqref{eq:U3d}, we get that
\begin{align}
\mathbf{u}_{\rm sd}^{\rm H\textrm{-}S}  &=- \dfrac{3 a^3U}{2h}
\left(\dfrac{\mathbf{I}}{\|\xb\|^2} -  2\dfrac{\xb\xb}{\|\xb\|^4} \right) \cdot \mathbf{p}\nonumber\\
&= \dfrac{ a^3A_1 M_0}{2hD}
\left(\dfrac{\mathbf{I}}{\|\xb\|^2} -  2\dfrac{\xb\xb}{\|\xb\|^4} \right) \cdot \mathbf{p}.
\label{eq:flowsourcedipole2d}
\end{align}
It is important to note that in unbounded $3D$ domains, the contribution of the source dipole was a higher order correction to the contribution of the force dipole. The situation is reversed in Hele-Shaw confinement: the source dipole decays as $1/\|\xb\|^2$ whereas the force dipole decays as $1/\|\xb\|^3$. Meanwhile, the force quadrupole decays as $1/\|\xb\|^4$. 
Ignoring the latter, the leading-order terms in the flow field created by a self-propelled phoretic particle placed horizontally in the mid-plane of the channel in the Hele-Shaw limit is computed by substituting Eqs.~\eqref{eq:flowforcedipole2d} and~\eqref{eq:flowsourcedipole2d} 
into Eq.~\eqref{eq:velfield3d},
\begin{align}
\mathbf{u}^{\rm H\textrm{-}S}&= \dfrac{ a^3A_1 M_0}{2hD} \left(\dfrac{\mathbf{I}}{\|\xb\|^2} -  2\dfrac{\xb\xb}{\|\xb\|^4} \right) \cdot \mathbf{p} \nonumber\\
 &+ \dfrac{5ha^2 \kappa A_1 M_1}{9D}  \left[ \grad \left(\dfrac{\mathbf{I}}{\|\xb\|^2} -  2\dfrac{\xb\xb}{\|\xb\|^4} \right)\right]:(3\pb\pb-\mathbf{I}).
 \label{eq:flow2d}
 \end{align}
The leading-order terms in the hydrodynamic and chemical signatures,~$\ub^{\rm H\textrm{-}S}(\xb)$ and $C^{\rm H\textrm{-}S}(\xb)$, of a self-propelled phoretic particle in the Hele-Shaw limit are summarized in table~\ref{tab:signature2d}.

\subsection{Hydrodynamic and phoretic interactions under weak confinement}

Under weak confinement $a\ll h$, the hydrodynamic and phoretic drifts of an individual particles -- which result from the chemical and hydrodynamic fields immediately around it -- can be determined as in the unbounded case. The hydrodynamic and phoretic drift velocities are therefore given by Eqs.~\eqref{eq:drift} and~\eqref{eq:chemdriftvel} respectively, where $\mathbf{u}_\infty$ and $\grad C_\infty$ are the velocity field and  concentration gradient created by other phoretic particles, and should now be evaluated for $\ub^{\rm H\textrm{-}S}(\xb)$ and $C^{\rm H\textrm{-}S}(\xb)$ in the Hele-Shaw limit obtained in Eqs.~\eqref{eq:chemical2d} and \eqref{eq:flow2d}.

Using Eqs.~\eqref{eq:chemdriftvel} and~\eqref{eq:chemical2d}, the phoretic drifts of a particle with orientation $\pb_0$ created by a particle located at relative position $\xb$ with orientation $\pb$ are given by
\begin{align}
\Ub^c_d&=-\dfrac{a^3 M_0A_1}{Dh}\left[\dfrac{\mathbf{I}}{\|\xb\|^2}-\dfrac{2\xb\xb}{\|\xb\|^4}\right]\cdot\pb, \\
 \Omegab^c_d&=-\frac{3a^2A_1M_1}{4Dh}\,\pb_0\times\left[\left(\dfrac{\mathbf{I}}{\|\xb\|^2}-\dfrac{2\xb\xb}{\|\xb\|^4}\right)\cdot\pb\right] ,
\end{align}
and respectively scale as $U_d^c \sim ({a^3}/{hL^2}) ({A_1M_0}/{D})$ and $\Omega_d^c \sim ({a^3}/{hL^2})({M_1A_1}/{aD})$.
Similarly, the velocity of translational drift due to hydrodynamic interactions is computed from Eqs.~\eqref{eq:drift} and~\eqref{eq:flow2d} 
\begin{align}
\label{eq:Uh_drift}
\Ub_d^h=&\frac{a^3A_1M_0}{2Dh}\left(\dfrac{\mathbf{I}}{\|\xb\|^2}-\frac{2\xb\xb}{\|\xb\|^4}\right)\cdot\mathbf{p} \nonumber\\
&+ \dfrac{5ha^2\kappa A_1M_1}{9D}\left[\grad\left(\dfrac{\mathbf{I}}{\|\xb\|^2}-\frac{2\xb\xb}{\|\xb\|^4}\right)\right]:(3\pb\pb-\mathbf{I}).
\end{align}
For spherical particles, the rotational hydrodynamic drift arises only from the vorticity field, as expressed explicitly in Eq.~\eqref{eq:drift}. Since the flow field in Eq.~\eqref{eq:flow2d} is potential, the vorticity field is identically zero, and therefore there is no hydrodynamic rotational drift in the Hele-Shaw limit.
A summary of the phoretic and hydrodynamic drift velocities in this limit is given in table~\ref{tab:drift2d}.

A direct comparison of the translational drift velocities $U_d^c$ and $U_d^h$ shows that the contributions of the hydrodynamic and chemical source dipoles follow the same scaling, while that due to the hydrodynamic force quadrupole is always subdominant; see table~\ref{tab:drift2d}. Comparing the velocities due to hydrodynamic and chemical source dipoles to that induced by the force dipole, three regimes can arise depending on the relative magnitude of $a/h$ versus $h/L$, or on the relative magnitude of the range of the interactions $L$ versus $h^2/a$:
 \begin{itemize}
 \item{\emph{For far-field interactions $L\gg h^2/a$,}}  the translational drift velocity $U_d$ is governed at the leading order by the hydrodynamic 
 and chemical source dipoles. The  influence of the force dipole is negligible.  
 \item{\emph{For short-range interactions $L\ll h^2/a$,}}  the translational drift due to the hydrodynamic force dipole is dominant. The chemical drift is negligible, and so is the role of the hydrodynamic source dipole.
\item{\emph{When $L\sim h^2/a$,}} the drifts induced by the hydrodynamic force dipole and source dipole and the chemical source dipole are all of the same order.
\end{itemize}
Effectively, if the particles' density is small enough (dilute systems) that the typical distance between particles satisfies $L\gg h^2/a$, then the leading-order interactions of the different particles can be written solely in terms of source dipoles. Higher particle densities require more care to account for the effect of the force dipole. In all regimes, one can readily verify that the rotational drift due to the chemical source dipole is always dominant. Starting from this insight, we next formulate equations of motion governing the far-field interaction of multiple phoretic janus particles in each of these regimes.

\begin{table*}[!t]
  \begin{center}    
         \begin{tabular}{l|ccc|cc}
   & \multicolumn{3}{c|}{\textbf{Hydrodynamic drift}} & \multicolumn{2}{c}{\textbf{Chemical drift}}\\[0.5ex] 
 &  force dipole & source dipole    & force quadrupole & source &  source dipole \\[0.5ex] \hline 
  & & & & & \\[-0.75ex]
translational drift $U_d$  & $\dfrac{h a^2}{L^3} \dfrac{A_1M_1}{D}$ & $\dfrac{a^3}{hL^2} \dfrac{A_1M_0}{D}$  
&
$\dfrac{a^3\alpha_3}{hL^4}$ 
& 
$-$ & $\dfrac{a^3}{hL^2} \dfrac{A_1M_0}{D}$  \\[2.5ex]
rotational drift $\Omega_d$ &$ -$&$-$  & $ -$   & 
$-$
& $\dfrac{1}{a}\dfrac{a^3}{hL^2} \dfrac{A_1M_1}{D}$ \\[0.5ex] 
        \end{tabular}
          \caption{Scaling laws of the chemical and hydrodynamic drift under confinement created by a phoretic particle at a distance $L$ from the particle of interest.}
  \label{tab:drift2d}
  \end{center}
\end{table*}



 
\section{Far-field interactions of confined auto-phoretic particles} 
\label{sec:2particles} 
Based on the results obtained in the previous section, we formulate a self-consistent description for the dynamics of $N$ autophoretic particles in weak confinement ($a\ll h$) and dilute suspensions ($h\ll L$). 

Let particle $j$ be located at $\mathbf{x}_j$  with orientation given by the unit vector $\mathbf{p}_j$ ($j=1,\ldots,N$). For simplicity, we consider that all particles are located on the midplane of the channel ($\delta=0$).  The translational motion of  particle $j$ is due (i) to its self-propulsion at speed $U$ as a result of its own chemical activity and mobility property, (ii) the hydrodynamic drift generated by the motion of its neighbors and (iii) the chemical drift resulting from the  particle's own mobility in response to the chemical activity of its neighbors. The particle's orientation $\pb_j$ also changes in response to these flow and chemical disturbances. 
Given $a\ll h$, the resulting equations of motion of particle $j$ are given at the leading order by the particle's behavior in unbounded flows
 \begin{align}
 \dot{\mathbf{x}}_j   & = U \mathbf{p}_j +  \mathbf{u}(\mathbf{x}_j) + \mu_c \nabla C ({\mathbf{x}_j}),  \label{eq:eom_general1}\\[2ex]
  \dot{\mathbf{p}}_j  & =  (\mathbf{I} - \mathbf{p}_j \mathbf{p}_j)\cdot \left[\mathbf{w}(\xb_j)\cdot\mathbf{p}_j  + \nu_c  \nabla C ({\mathbf{x}_j}) \right].
 \label{eq:eom_general2}
 \end{align}
Here, $U$ is the self-propulsion velocity given in Eq.~\eqref{eq:U3d},  $\mu_c$ and $\nu_c$ are the chemical translational mobility coefficients, and $\mathbf{w}=(\nabla \mathbf{u}-\nabla \mathbf{u}^T)/2$  is anti-symmetric (vorticity) components of the local velocity gradient.
In Section~\ref{sec:unbounded}, we obtained that, for a spherical Janus particle, $U$, $\mu_c$ and $\nu_c$ are given by Eq.~\eqref{eq:chemdriftvel}, which we rewrite for convenience,
\begin{align}
U=-\frac{A_1M_0}{3D},\qquad\mu_c=-M_0,\qquad \nu_c=-\frac{3M_1}{4a}.
\end{align} 
 
In section~\ref{sec:confined}, we found that the rotational drift is always dominated by the chemical component. Therefore, in the following, we will neglect the hydrodynamic term in the orientation equation.
We also obtained expressions for the local hydrodynamic and chemical fields generated by other particles. Namely, the chemical concentration $C$ is given by Eq.~\eqref{eq:chemical2d} and the hydrodynamic flow field $\ub$ by Eq.~\eqref{eq:flow2d}. Here, we drop the H-S superscript with the understanding that all quantities are in  Hele-Shaw confinement. Put together, we get that, at the leading order,
\begin{align}
C(\xb)&=\frac{a^3A_1}{Dh}\sum_{k}\frac{\pb_k\cdot(\xb - \xb_k)}{\|\xb - \xb_k\|^2}
\\
  \nabla C(\xb)&=-\frac{a^3A_1}{Dh}\sum_{k} \mathbf{G}(\xb,\xb_k)\cdot\pb_k 
\end{align}
and
\begin{align}
\ub&(\xb)
 =-\dfrac{a^3A_1M_0}{2Dh} \sum_k \mathbf{G}(\xb,\xb_k) \cdot \mathbf{p}_k \nonumber\\
&  -
\dfrac{5ha^2\kappa A_1M_1}{9}   \sum_k
 \nabla \mathbf{G}(\xb,\xb_k):(3\pb_k\pb_k-\mathbf{I}),
\label{eq:flow}
\end{align}
with 
\begin{equation}
\mathbf{G}(\xb,\xb_k)=  \dfrac{2(\xb - \xb_k)(\xb - \xb_k)}{\|\xb - \xb_k\|^4} -\dfrac{\mathbf{I}}{\|\xb - \xb_k\|^2} \cdot
\end{equation}
We evaluate the above expressions at particle $j$ and substitute the result in Eqs.~\eqref{eq:eom_general1}-\eqref{eq:eom_general2}, noting that the second term in the hydrodynamic flow field, Eq.~\eqref{eq:flow}, should only be included for particles when $|(\xb_j-\xb_k)|\lessapprox h^2/a$. For  dilute suspensions, where the volume fraction of particles is such that $|(\xb_j-\xb_k)|\gg h^2/a$, the contribution of the force dipole is subdominant. In this case, we note that the hydrodynamic and chemical translational drift on particle $j$ take the exact same form given by $\ub(\xb_j)$ and $-M_0\grad C(\xb_j)=-2\ub(\xb_j)$, but act in opposite directions, with the latter being dominant (and exactly twice as large). 
The equations of motion for the phoretic particles then simplify to
 \begin{align}
 \dot{\mathbf{x}}_j   & = -\frac{A_1M_0}{3D} \mathbf{p}_j +  \dfrac{a^3A_1M_0}{2Dh} \sum_{k\neq j} \mathbf{G}(\xb_j,\xb_k)\cdot \mathbf{p}_k , \label{eq:eom_final1}\\
  \dot{\mathbf{p}}_j  & =   
\frac{3a^2M_1A_1}{4Dh}  (\mathbf{I} - \mathbf{p}_j \mathbf{p}_j) \sum_{k\neq_j}\mathbf{G}(\xb_j,\xb_k)\cdot\pb_k	 .
 \label{eq:eom_final2}
 \end{align}
We note that these equations take a particularly simple form: they are equivalent to the interaction of hydrodynamic {source} dipoles with a reversed hydrodynamic drift, leading to a negative effective hydrodynamic mobility coefficient. That is to say, particles tend to drift in the opposite direction of the local flow due to the dominance of the chemical drift.

It is convenient to rewrite Eqs.~\eqref{eq:eom_final1}--\eqref{eq:eom_final2} in non-dimensional form. To this end, we assume without any loss of generality that $A_1M_0>0$ so that the direction of self-propulsion is $\mathbf{q}=-\pb$ (the case $A_1M_0<0$ could be treated similarly); we then use the characteristic length scale $a$ and a characteristic time scale $3Da/A_1 M_0$ based on the nominal swimming velocity. Equations~\eqref{eq:eom_final1}--\eqref{eq:eom_final2} become
  \begin{align}
 \dot{\mathbf{x}}_j   & =  \mathbf{q}_j +{\mu}  \sum_{k\neq j} \mathbf{G}(\xb_j,\xb_k) \cdot \mathbf{q}_k, \label{eq:eom_dimensionless}\\
  \dot{\mathbf{q}}_j  & =   
{\nu} (\mathbf{I} - \mathbf{q}_j \mathbf{q}_j) \sum_{k\neq j}\mathbf{G}(\xb_j,\xb_k)\cdot\mathbf{q}_k	.
 \label{eq:eom_dimensionlessb}
 \end{align}
where $\mu = -3a/2h$ and $\nu = 9aM_1/4hM_0$ are the translational and rotational mobility, respectively,  with the understanding that all variables in Eq.~\eqref{eq:eom_dimensionless}--\eqref{eq:eom_dimensionlessb} are non-dimensional. From a chemical point of view, the sign of $\nu$ is directly related to the reorientation of the particle in the direction of or opposed to its chemical drift. 

It should be noted here that the mobility coefficient  $\mu = - 3a/2h$ is negative. That is, in contrast with standard micro-swimmers~\cite{Desreumaux2012,Tsang2014,Tsang2015,Tsang2016}, weakly-confined phoretic particles drift in the opposite direction to the local flow induced by other particles, instead following their chemical drift. This, in turn, is the result of phoretic and hydrodynamics interactions having the same dependence but opposite behaviour in this Hele-Shaw limit.

\begin{figure*}[!t]
\centerline{\includegraphics[scale=1]{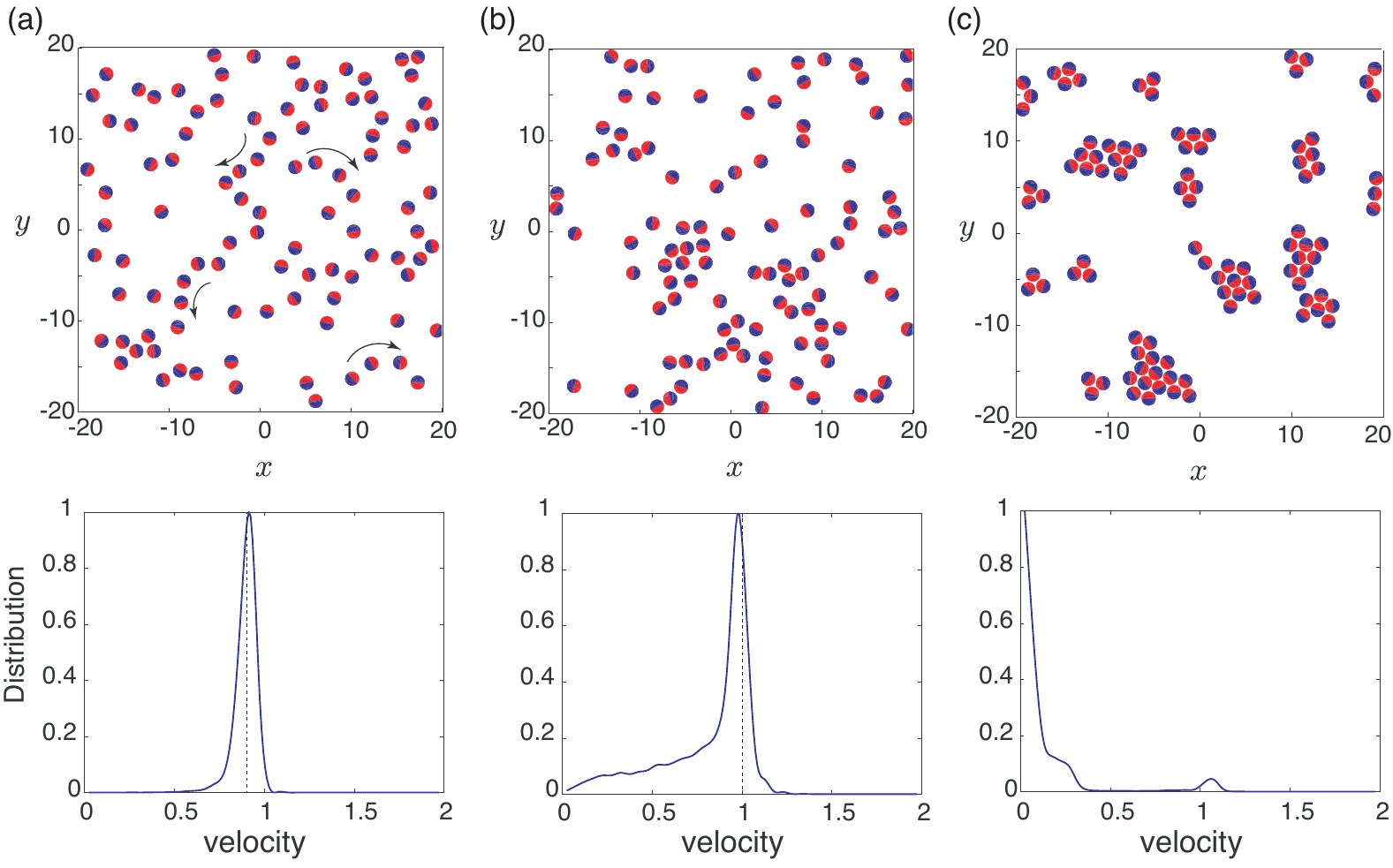}}
\caption[]{Collective behavior of auto-phoretic particles in doubly-periodic domain exhibiting (a) chain-like formation and swirling behavior for $\nu = 5$;
(b) advection and steric interaction for $\nu = 0$; and (c)  aggregation for $\nu = -5$. Snapshots are shown at $t = 1750$, $1000$ and $15$, respectively. {The unit vector $\mathbf{p}$ is directed from red to blue, as indicated in Figure 1 (color online)}. Parameter values are set to $N =100$, $U = 1$, $a = 1$,  $\mu = -0.3$. The domain size is $D = 19.5$. Bottom row shows the probability distributions of the particle velocities in the three cases. In all cases, the average velocity is smaller than the self-propelled velocity of an individual particle $U =1$.}
	\label{fig:behavior}
\end{figure*}

\section{Collective dynamics and pair interactions}
\label{sec:suspensions}
In the previous section, the leading order dynamics for phoretic particles in weak Hele-Shaw confinement was shown to take the form of interacting potential dipoles, similarly to other categories of confined micro-swimmers but for a reverse translational mobility.

\subsection{Suspension dynamics}
To analyze the interplay between this negative translational motility and the orientational dynamics within a suspension of phoretic particles, we numerically solve Eqs.~\eqref{eq:eom_dimensionless}--\eqref{eq:eom_dimensionlessb} for a population of particles in a doubly-periodic domain. We account for the doubly-infinite system of images using the Weiestrass-zeta function~\citep{Tsang2013,Tsang2014,Coquereaux1990}. To avoid collision, we introduce a Lennard-Jones repulsion potential to the translational equation, Eq.~\eqref{eq:eom_dimensionless}~\citep{Tsang2014}. The resulting steric  forces act locally and decay rapidly such that they do not affect the long-range chemical and hydrodynamic interactions among the particles. We use a standard time stepping algorithm to solve for the evolution of a population of $N = 100$ particles that are initially spatially distributed at random orientations in a doubly-periodic square domain of size $D/a = 19.5$~\citep{Tsang2014}. The aspect ratio is fixed to $h/a=5$ so that the translational mobility coefficient is $\mu = -0.3$ and the orientational coefficient is varied, $\nu \in [-5,5]$. 

Three distinct types of global behaviors emerge depending on the sign of the rotational mobility $\nu$: (i) a swirling behavior where particles form transient chains that emerge, break, and rearrange elsewhere for $\nu>0$, (ii) random particle motions for $\nu = 0$, and (iii) aggregation and clustering for $\nu<0$. Representative simulations are shown in Fig.~\ref{fig:behavior}. These collective phenomena are reminiscent to those observed in the motion of hydrodynamic dipoles~\citep{Tsang2014,Kanso2014}, but bear distinctive features due to the negative translational motility, as discussed next.

\begin{figure}[!t]
\centerline{\includegraphics[scale=1]{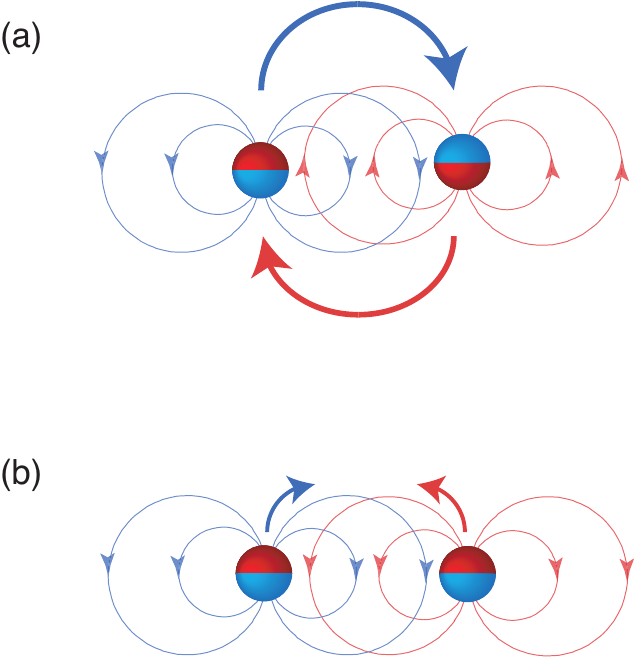}}
\caption[]{Schematic depiction of the orientation interaction of two phoretic particles: (a) for $\nu>0$, a particle reorients to align with the dipolar flow field created by the other particle, leading the two particles to ``tail gate" each other; (b) for {$\nu <0$}, it reorients opposite to the dipolar field created by the other particle, leading the two particles to aggregate head on.}
	\label{fig:reorient}
\end{figure}

The large-scale phenomena are dictated by the orientational dynamics of the particles. The swirling phenomenon can be readily understood by examining the orientational interaction of two particles, as depicted schematically in Fig.~\ref{fig:reorient}(a).  For $\nu>0$, a particle aligns with the drift created by another particle. That is to say, it reorients towards the tangent to the streamlines of the potential dipole created by the nearby particle, causing it naturally to follow that particle. As a result, particles of positive $\nu$ tend to form chains. However, given the negative motility coefficient $\mu<0$, as particles align into such chain-like structures, their forward translational motion is slowed down by the dipolar flow field of their neighbors, resulting in a decrease in the particles' velocities, as evidenced from the probability distribution function in Fig.~\ref{fig:behavior}(a).

When the orientational interactions are suppressed ($\nu =0$), particles do not change orientation, except to avoid collisions when other particles are sufficiently close to trigger steric interactions. In other words, the long-range interactions among particles can at most slow down the particle translational velocities due to the negative motility coefficient, without introducing bias in the particles' orientation and position relative to each other. The reduction in the translational velocity depends on the particles' location, which is initially random and remains random under subsequent interactions between particles, see Fig.~\ref{fig:behavior}(b). As a result, the translational velocities follow a somewhat broad-band distribution of values below $U=1$ (i.e. their self-propulsion velocity), as depicted in the bottom row of Fig.~\ref{fig:behavior}(b).

The clustering behavior observed for $\nu<0$ can be explained by recalling that, in that case, particles reorient in the opposite direction of the local drift resulting from the chemical and hydrodynamic interactions with the other particles. Therefore, a particle travels in the opposite direction to the streamlines created by a nearby particle, as illustrated in Fig.~\ref{fig:reorient}(b), which leads to aggregation. The particles begin to aggregate at a relatively short time scale, as indicated by the time of the snapshot in Fig.~\ref{fig:behavior}(c), and lead to the formation of clusters. Some of these clusters are unstable while others form stable aggregates that attract each other and slow down considerably as they coalesce to form larger aggregates. The velocity distribution function on the bottom row of Fig.~\ref{fig:behavior}(c) has a strong peak around zero velocity because most particles are attached to stable clusters, with a few particles moving at unit speed.

\subsection{Pair interactions and stability}

The orientational interactions underlying these large-scale phenomena can be examined analytically in more details in the special case of pairs of phoretic particles. In particular, the observations that
for $\nu>0$, particles tend to follow each other and form quasi-stable chainlike structures whereas for $\nu<0$, particles tend to collide head-on and form clusters are rooted in two-particle interactions: for $\nu<0$,  chainlike structures are unstable while side-by-side motions that lead to head-on collisions are unstable for $\nu>0$, as discussed next.

\begin{figure}[!t]
\centerline{\includegraphics[scale=1]{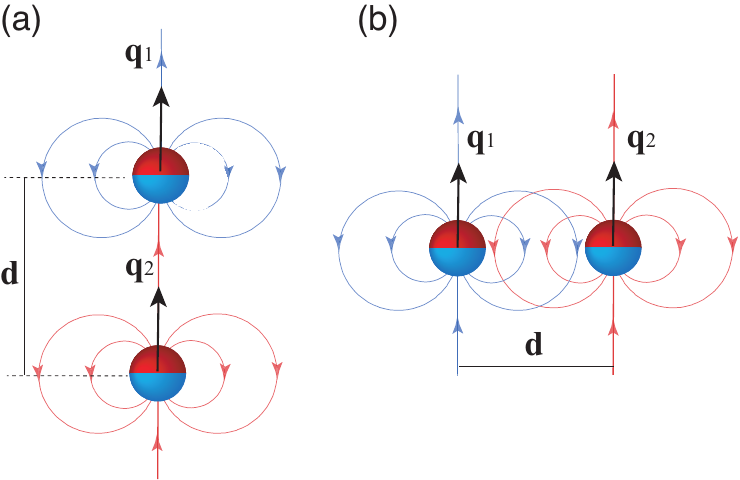}}
\caption[]{Relative equilibria of pairs of particles in (a) \textit{follower} configuration and (b) \textit{side-by-side} configuration, both at a constant separation distance $d_o$. The follower configuration is unstable for $\nu<0$ while the side-by-side configuration is unstable for $\nu>0$.}
	\label{fig:equilibria}
\end{figure}

First, we rewrite Eqs.~\eqref{eq:eom_dimensionless}--\eqref{eq:eom_dimensionlessb} for two particles to get that
\begin{align}
\dot{\xb}_1&=\mathbf{q}_1+\frac{\mu}{d^2}\left(\frac{2\mathbf{d}\mathbf{d}-\mathbf{I}}{d^2}\right)\cdot\mathbf{q}_2,\\
\dot{\xb}_2&=\mathbf{q}_2+\frac{\mu}{d^2}\left(\frac{2\mathbf{d}\mathbf{d}}{d^2}-\mathbf{I}\right)\cdot\mathbf{q}_1,\\
\dot{\mathbf{q}}_1&=\frac{\nu}{d^2}(\mathbf{1}-\mathbf{q}_1\mathbf{q}_1)\cdot\left(\frac{2\mathbf{d}\mathbf{d}}{d^2}-\mathbf{I}\right)\cdot\mathbf{q}_2,\\
\dot{\mathbf{q}}_2&=\frac{\nu}{d^2}(\mathbf{1}-\mathbf{q}_2\mathbf{q}_2)\cdot\left(\frac{2\mathbf{d}\mathbf{d}}{d^2}-\mathbf{I}\right)\cdot\mathbf{q}_1,\label{eq:p2}
\end{align}
where $\mathbf{d}=\xb_2-\xb_1$ is the relative position vector and $d = \| \mathbf{d} \|$ is the relative distance between the two particles. 
We reformulate these equations in terms of $\mathbf{d}$ and the  relative orientation angles $\alpha_j$ of $\mathbf{q}_j$ ($j=1,2$) with respect to $\mathbf{d}$. To this end, we introduce the unit vector  $\eb= \mathbf{d}/d \equiv (\cos\theta,\sin\theta)$, where $\theta$ is the orientation of $\mathbf{d}$ in the fixed inertial frame. The relative angles $\alpha_j$ of $\mathbf{q}_j$ with respect to $\eb$ satisfy the identities $\eb\cdot\mathbf{q}_j=\cos\alpha_j$ and $\eb\times\mathbf{q}_j=\sin\alpha_j$. The global translational velocity of the two particles can be omitted here due to the fact that the system is invariant under translational symmetry. Expressing Eq.~\eqref{eq:p2} in terms of $(d,\theta,\alpha_1,\alpha_2)$, we get that
\begin{align}
\dot{d}&=\left(1-\frac{\mu}{d^2}\right)(\cos\alpha_2 - \cos\alpha_1),\\
\dot\theta  &=\left(\frac{1}{d}+\frac{\mu}{d^3}\right)(\sin\alpha_2-\sin\alpha_1), \\
\dot\alpha_1& =\dot\alpha_2=-\frac{\nu}{d^2}\sin(\alpha_1+\alpha_2) \nonumber \\ &\qquad \qquad -\left(\frac{1}{d}+\frac{\mu}{d^3}\right)(\sin\alpha_2-\sin\alpha_1).
\end{align}
This leads to the surprising result that the relative orientation $\alpha_2 - \alpha_1$ of the two swimmers is a constant of motion. Further, $\theta$ does not influence $(d,\alpha_1,\alpha_2)$ because of the system is invariant under rotational symmetry; we could thus solve for $(d,\alpha_1,\alpha_2)$ independently. It is more convenient to introduce
$\delta =( \alpha_2 - \alpha_1)/2$ and $\gamma = (\alpha_1 + \alpha_2)/2$, such that 
\begin{align}
\dot{d}&=\left(1-\frac{\mu}{d^2}\right) \sin\gamma\sin\delta, \\
\dot\delta &= 0, \\
\dot\gamma& =-\frac{\nu}{d^2}\sin 2 \gamma - 2\left(\frac{1}{d}+\frac{\mu}{d^3}\right)\cos\gamma\sin\delta.
\label{eq:shape}
\end{align}
There are two configurations that lead to relative equilibria of this system of equations ($\dot\delta=\dot\gamma=0$): (i) \emph{Follower configuration:} $\delta=0$, $\gamma=0$, and (ii) \emph{Side-by-side configuration}: $\delta=0$, $\gamma=\pm\pi/2$, as depicted in Fig.~\ref{fig:equilibria}.

In the \emph{follower} configuration, both particles have the same orientation aligned with their relative distance $\mathbf{e}$. The solution corresponds to a translation of the two particles at the same velocity $(1+2\mu/d_o^2)\eb$, where $d_o$ is constant. Linearizing Eq.~\eqref{eq:shape} around the equilibrium $(d_o, 0, 0)$ provides at leading order,
\begin{equation}
\dot{d} = 0, \quad \dot{\delta} = 0, \quad \dot{\gamma} = -2\frac{\nu}{d_o^2} \gamma - 2\left(\frac{1}{d_o}+\frac{\mu}{d_o^3}\right)\delta,
\end{equation}
and this equilibrium is linearly unstable for $\nu<0$.  For $\nu\geq 0$, the system is neutrally stable and weakly nonlinear analysis should be performed to analyze its stability. These findings are consistent with the observations that chainlike structures, which are reminiscent of the follower configuration, are not observed for $\nu<0$. Physically-speaking, the follower configuration is unstable for $\nu<0$ because particles tend to align in the opposite direction to the ambient flow. Therefore, a slight perturbation away from the equilibrium, say of the follower particle, would cause it to move opposite to the streamlines of the leader particle, which drives it further away from the equilibrium.

In the \emph{Side-by-side} configuration, both particles are parallel to each other and perpendicular to $\mathbf{e}$. They undergo a translational motion with velocity $(1-\mu/d_o^2)\eb$, where $d_o$ is constant. Linearizing around $(d_o,0, \pm\pi/2)$ provides at leading order,
\begin{equation}
\dot{d} = \pm \left(1-\frac{\mu}{d_o^2}\right)\delta, \qquad \dot{\delta} = 0, \qquad \dot{\gamma} =  2\frac{\nu}{d_o^2} \gamma,
\end{equation}
 and this equilibrium is linearly unstable for $\nu>0$ and neutrally stable otherwise. These results are consistent with the intuitive analysis presented in Fig.~\ref{fig:reorient}. Side-by-side particles pointing in the same direction are unstable for $\nu>0$ because they will want to reorient in the direction of the dipolar field of the other particle. 

Taken together, the results from the numerical simulations for a population of weakly-confined particles and the linear stability analysis for a pair of particles indicate that the collective behavior is dominated by the orientational dynamics and by the orientational mobility coefficient $\nu$. The negative translational mobility coefficient slows down the particles but does not affect the modes of collective behavior.

\section{Conclusions}\label{sec:conclusions}
 
In this work, a first-principle approach was presented for deriving the equations governing the interactions of a population of auto-phoretic Janus particles under weak Hele-Shaw confinement. 
The goal was to analyse the effects of confinement on the interactions of many phoretic particles, and in particular on the relative weight of the hydrodynamic and chemical (phoretic) couplings.
Both effects take the same form within this particular limit, but act in opposite directions, with the magnitude of phoretic interactions being exactly twice as large, and therefore driving the collective dynamics of weakly-confined Janus particles. Yet, hydrodynamic interactions are not negligible, and in fact account for a reduction by $50\%$ in the effective translational and rotational drifts when compared to pure phoretic interactions. This analysis further provides a detailed insight on the relative weight of hydrodynamics and phoretic drift by obtaining precise and comparative scalings for the two routes of interactions in unbounded and confined geometries.
 
In the Hele-Shaw limit considered, the leading order dynamics takes the form of interacting potential dipoles, similarly to other categories of confined micro-swimmers\citep{Tsang2014,Tsang2015,Tsang2016} but with a reverse translational mobility due to the phoretic coupling. 
The reverse translational motility slows down the particles but does not affect the emergent collective modes, which are governed by the sign of the orientational mobility coefficient $\nu$. 
Particles that align with the drift created by the other particles ($\nu>0$) exhibit global swirling and chaotic-like behavior, while particles that align opposite to the induced drift ($\nu<0$) tend to aggregate and form stationary clusters. 

These collective modes were previously analyzed, albeit for micro-swimmers with positive translational mobility coefficient, showing that the transition from swirling  to clustering and aggregation as  $\nu$  decreases from positive to negative occurs systematically over the phase space consisting of $\nu$ and the particles' area fraction $\Phi$ (with $\Phi$ being the ratio of the area of of all particles to the area of the doubly-periodic domain) taken to lie in the dilute to semi-dilute range $\Phi \in [0.1, 0.3]$~\citep{Tsang2014}. 
Therefore, we expect the global swirling and aggregation modes in the weakly-confined auto-phoretic particles, and the transition between them as $\nu$ decreases, 
to be robust to changes in the number of particles in this range of $\Phi$. 
Further, these global modes were shown to be robust to rotational Brownian noise for a range of rotational diffusion coefficients with P\'{e}clet numbers of order 1~\citep{Tsang2014}.
We thus expect rotational diffusion in this range of P\'{e}clet numbers to have a small effect on the collective modes of weakly-confined auto-phoretic particles.
{In fact, the framework presented here purposely neglects the influence of thermal fluctuations and the Brownian nature of the dynamics of small Janus colloids, as our focus was on understanding the role of confinement in screening and tuning each route of deterministic interactions between chemically-active colloids. Future studies will address in details how such screening is potentially influenced by stochastic fluctuations in the particles' dynamics.}

These findings, albeit in the context of a simplified model, may have profound implications on understanding and controlling the collective behavior of active films by auto-phoretic particles.
They demonstrate that, in weak Hele-Shaw confinement, the emergent phase is controllable by the surface properties of the individual Janus particles. 
The surface chemistry dictate the ability of a Janus particle to drive surface slip from local concentration gradients, which, in turn, dictates the sign and value of $\nu$. 
Therefore, the mobility and chemical activity ($M$ and $A$) can be viewed as control parameters to systematically and predictably engineer active films with distinct emergent properties, from spontaneous large-scale swirling motions to stationary clusters and aggregates.

 \paragraph*{Acknowledgement.} This work was partially supported by a visiting scholar position from the Laboratoire d'Hydrodynamique LadHyX, D\'epartement de M\'ecanique, Ecole Polytechnique, and by the National Science Foundation via the NSF INSPIRE (grant 16-08744 to E.K.). This work was also supported by the European Research Council (ERC) under the European Union's Horizon 2020 research and innovation program (grant agreement 714027 to S.M.).



\begin{thebibliography}{10}

\bibitem{lauga2009}
E.~Lauga and T.~R. Powers.
\newblock The hydrodynamics of swimming microorganisms.
\newblock {\em Reports on Progress in Physics}, 72(9):096601, 2009.

\bibitem{brennen1977}
C.~Brennen and H.~Winnet.
\newblock Fluid mechanics of propulsion by cilia and flagella.
\newblock {\em Annu. Rev. Fluid Mech.}, 9:339--398, 1977.

\bibitem{lauga2016}
E.~Lauga.
\newblock Bacterial hydrodynamics.
\newblock {\em Annu. Rev. Fluid Mech.}, 48:105--130, 2016.

\bibitem{dreyfus2005}
R.~Dreyfus, J.~Baudry, M.~L. Roper, M.~Fermigier, H.~A. Stone, and J.~Bibette.
\newblock Microscopic artificial swimmers.
\newblock {\em Nature}, 437:862--865, 2005.

\bibitem{bricard2013}
Antoine Bricard, Jean-Baptiste Caussin, Nicolas Desreumaux, Olivier Dauchot,
  and Denis Bartolo.
\newblock Emergence of macroscopic directed motion in populations of motile
  colloids.
\newblock {\em Nature}, 503(7474):95--98, 2013.

\bibitem{wang2012}
W.~Wang, L.A. Castro, M.~Hoyos, and T.E. Mallouk.
\newblock Autonomous motion of metallic microrods propelled by ultrasound.
\newblock {\em ACS Nano}, 6(7):6122--6132, 2012.

\bibitem{anderson1989}
J.~L. Anderson.
\newblock Colloid transport by interfacial forces.
\newblock {\em Annu. Rev. Fluid Mech.}, 21:61--99, 1989.

\bibitem{paxton2004}
W.~F. Paxton, K.~C. Kistler, C.~C. Olmeda, A.~Sen, S.~K.~St. Angelo, Y.~Cao,
  T.~E. Mallouk, P.~E. Lammert, and V.~H. Crespi.
\newblock {Catalytic Nanomotors: Autonomous Movement of Striped Nanorods}.
\newblock {\em J. Am. Chem. Soc.}, 126(41):13424--13431, 2004.

\bibitem{howse2007}
J.~R. Howse, R.~A.~L. Jones, A.~J. Ryan, T.~Gough, R.~Vafabakhsh, and
  R.~Golestanian.
\newblock {Self-Motile Colloidal Particles: From Directed Propulsion to Random
  Walk}.
\newblock {\em Phys. Rev. Lett.}, 99(4):048102, 2007.

\bibitem{theurkauff2012}
I.~Theurkauff, C.~Cottin-Bizonne, J.~Palacci, C.~Ybert, and L.~Bocquet.
\newblock Dynamic clustering in active colloidal suspensions with chemical
  signaling.
\newblock {\em Phys. Rev. Lett.}, 108:268303, 2012.

\bibitem{palacci2013}
J.~Palacci, S.~Sacanna, A.~P. Steinberg, D.~J. Pine, and P.~M. Chaikin.
\newblock Living crystals of light-activated colloidal surfers.
\newblock {\em Science}, 339:936--940, 2013.

\bibitem{brown2014}
A.~Brown and W.~Poon.
\newblock Ionic effects in self-propelled pt-coated janus swimmers.
\newblock {\em Soft Matter}, 10:4016--4027, 2014.

\bibitem{ibele2009}
M.~Ibele, T.~E. Mallouk, and A.~Sen.
\newblock Schooling behavior of light-powered autonomous micromotors in water.
\newblock {\em Angew. Chem. Int. Ed.}, 48(18):3308--3312, 2009.

\bibitem{izri2014}
Z.~Izri, M.~N. van~der Linden, S.~Michelin, and O.~Dauchot.
\newblock Self-propulsion of pure water droplets by spontaneous
  marangoni-stress-driven motion.
\newblock {\em Phys. Rev. Let.}, 113:248302, 2014.

\bibitem{herminghaus2014}
S.~Herminghaus, C.~C. Maass, C.~Kr{\"u}ger, S.~Thutupalli, L.~Goehring, and
  C.~Bahr.
\newblock Interfacial mechanisms in active emulsions.
\newblock {\em Soft Matter}, 10:7008, 2014.

\bibitem{buttinoni2013}
I.~Buttinoni, J.~Bialk{\'e}, F.~K\"ummel, H.~L\"owen, C.~Bechinger, and
  T.~Speck.
\newblock Dynamical clustering and phase separation in suspensions of
  self-propelled colloidal particles.
\newblock {\em Phys. Rev. Lett.}, 110:238301, 2013.

\bibitem{colberg2017}
P.~Colberg and R.~Kapral.
\newblock Many-body dynamics of chemically propelled nanomotors.
\newblock {\em J. Chem. Phys.}, 147:064910, 2017.

\bibitem{ilien2017}
P.~Ilien, R.~Golestanian, and A.~Sen.
\newblock "fuelled" motion: phoretic motility and collective behaviour of
  active colloids.
\newblock {\em Chem. Soc. Rev.}, 46:5508--5518, 2017.

\bibitem{ginot2018}
F.~Ginot, I.~Theurkauff, F.~Detcheverry, C.~Ybert, and C.~Cottin-Bizonne.
\newblock Aggregation-fragmentation and individual dynamics of active clusters.
\newblock {\em Nat. Comm.}, 9:696, 2018.

\bibitem{varma2018}
A.~Varma, T.~D. Montenegro-Johnson, and S.~Michelin.
\newblock Clustering-induced self-propulsion of isotropic phoretic particles.
\newblock {\em Soft Matter}, 14:7155, 2018.

\bibitem{ke2010}
H.~Ke, S.~Ye, R.~L. Carroll, and K.~Showalter.
\newblock Motion analysis of self-propelled pt-silica particles in hydrogen
  peroxide solutions.
\newblock {\em J. Phys. Chem. A}, 114:5462--5467, 2010.

\bibitem{valadares2010}
L.~F. Valadares, Y.-G. Tao, N.~S. Zacharia, V.~Kitaev, F.~Galembeck, R.~Kapral,
  and G.~A. Ozin.
\newblock Catalytic nanomotors: self-propelled sphere dimers.
\newblock {\em Small}, 6:565--572, 2010.

\bibitem{ebbens2011}
S.~J. Ebbens and J.~R. Howse.
\newblock Direct observation of the direction of motion for spherical catalytic
  swimmers.
\newblock {\em Langmuir}, 27:12293--12296, 2011.

\bibitem{volpe2011}
G.~Volpe, I.~Buttinoni, D.~Vogt, H.-J. K\"ummerer, and C.~Bechinger.
\newblock Microswimmers in patterned environments.
\newblock {\em Soft Matter}, 7:8810--8815, 2011.

\bibitem{baraban2013}
L.~Baraban, R.~Streubel, D.~Makarov, L.~Han, D.~Karnaushenko, O.~G. Schmidt,
  and G.~Cuniberti.
\newblock Fuel-free locomotion of janus motors: magnetically-induced
  thermophoresis.
\newblock {\em ACS Nano}, 7:1360--1367, 2013.

\bibitem{davieswykes2017}
M.~S.~Davies Wykes, X.~Zhong, J.~Tong, T.~Adachi, Y.~Liu, M.~D. Ward, M.~J.
  Shelley, and J.~Zhang.
\newblock Guiding microscale swimmers using teardrop-shaped posts.
\newblock {\em Soft Matter}, 13:4681, 2017.

\bibitem{wang2006}
Y.~Wang, R.~M. Hernandez, D.~J.~Bartlett Jr., J.~M. Bingham, T.~R. Kline,
  A.~Sen, and T.~E. Mallouk.
\newblock Bipolar electrochemical mechanism for the propulsion of catalytic
  nanomotors in hydrogen peroxide solutions.
\newblock {\em Langmuir}, 22:10451--10456, 2006.

\bibitem{moran2017}
J.~L. Moran and J.~D. Posner.
\newblock Phoretic self-propulsion.
\newblock {\em Annu. Rev. Fluid Mech.}, 49:511--540, 2017.

\bibitem{ibrahim2017}
Y.~Ibrahim, R.~Golestanian, and T.~B. Liverpool.
\newblock Multiple phoretic mechanisms in the self-propulsion of a pt-insulator
  janus swimmers.
\newblock {\em J. Fluid Mech.}, 828:318, 2017.

\bibitem{saha2014}
S.~Saha, R.~Golestanian, and S.~Ramaswamy.
\newblock Clusters, asters, and collective oscillations in chemotactic
  colloids.
\newblock {\em Phys. Rev. E}, 89:062316, 2014.

\bibitem{tatuleacodrean2018}
M.~Tatulea-Codrean and E.~Lauga.
\newblock Artificial chemotaxis of phoretic swimmers: instantaneous and
  long-time behaviour.
\newblock {\em J. Fluid Mech.}, (to appear), 2018.
\newblock arxiv:1809.02374v1.

\bibitem{soto2014}
R.~Soto and R.~Golestanian.
\newblock Self-assembly of catalytically active colloidal molecules: tailoring
  activity through surface chemistry.
\newblock {\em Phys. Rev. Lett.}, 112:068301, 2014.

\bibitem{liebchen2015}
B.~Liebchen, D.~Marenduzzo, I.~Pagonabarraga, and M.~E. Cates.
\newblock Clustering and pattern formation in chemorepulsive active colloids.
\newblock {\em Phys. Rev. Lett.}, 115:258301, 2015.

\bibitem{liebchen2017}
B.~Liebchen, D.~Marenduzzo, and M.~E. Cates.
\newblock Phoretic interactions generically induce dynamic clusters and wave
  patterns in active colloids.
\newblock {\em Phys. Rev. Lett.}, 118:268001, 2017.

\bibitem{huang2017}
M.-J. Huang, J.~Schofield, and R.~Kapral.
\newblock Chemotactic and hydrodynamic effects on collective dynamics of
  self-diffusiophoretic janus motors.
\newblock {\em New J. Phys.}, 19:125003, 2017.

\bibitem{palacci2010}
J.~Palacci, C.~Cottin-Bizonne, C.~Ybert, and L.~Bocquet.
\newblock {Sedimentation and Effective Temperature of Active Colloidal
  Suspensions}.
\newblock {\em Phys. Rev. Lett.}, 105(8):088304, 2010.

\bibitem{kruger2016}
C.~Kr{\"u}ger, C.~Bahr, S.~Herminghaus, and C.~C. Maass.
\newblock Dimensionality matters in the collective behaviour of active
  emulsions.
\newblock {\em Eur. Phys. J. E}, 39:64--72, 2016.

\bibitem{golestanian2007}
R.~Golestanian, T.~B. Liverpool, and A.~Ajdari.
\newblock {Designing phoretic micro- and nano-swimmers}.
\newblock {\em New J. Phys.}, 9:126, 2007.

\bibitem{julicher2009}
F~J{\"u}licher and J~Prost.
\newblock {Generic theory of colloidal transport}.
\newblock {\em Eur. Phys. J. E}, 29(1):27--36, 2009.

\bibitem{michelin2014}
S.~Michelin and E.~Lauga.
\newblock Phoretic self-propulsion at finite p{\'e}clet numbers.
\newblock {\em J. Fluid Mech.}, 747:572--604, 2014.

\bibitem{uspal2014}
W.~E. Uspal, M.~N. Popescu, S.~Dietrich, and M.~Tasinkevych.
\newblock Self-propulsion of a catalytically active particle near a planar
  wall: from reflection to sliding and hovering.
\newblock {\em Soft Matter}, 11:434--438, 2015.

\bibitem{uspal2016}
W.~E. Uspal, M.~N. Popescu, S.~Dietrich, and M.~Tasinkevych.
\newblock Guiding catalytically active particles with chemically patterned
  surfaces.
\newblock {\em Phys. Rev. Lett.}, 117:048002, 2016.

\bibitem{mozaffari2016}
A.~Mozaffari, N.~Sharifi-Mood, J.~Koplik, and C.~Maldarelli.
\newblock Self-diffusiophoretic colloidal propulsion near a solid boundary.
\newblock {\em Phys. Fluids}, 28:053107, 2016.

\bibitem{malgaretti2018}
P.~Malgaretti, M.~N. Popescu, and S.~Dietrich.
\newblock Self-diffusiophoresis induced by fluid interfaces.
\newblock {\em Soft Matter}, 14:1375, 2018.

\bibitem{das2015}
S.~Das, A.~Garg, A.~I. Campbell, J.~Howse, A.~Sen, D.l Velegol, R.~Golestanian,
  and S.~J. Ebbens.
\newblock Boundaries can steer active janus spheres.
\newblock {\em Nat. Comm.}, 6:8999, 2015.

\bibitem{simmchen2016}
J.~Simmchen, J.~Katuri, W.~E. Uspal, M.~N. Popescu, M.~Tasinkevych, and
  S.~S\'anchez.
\newblock Topographical pathways guide chemical microswimmers.
\newblock {\em Nat. Comm.}, 7:10598, 2016.

\bibitem{dominguez2016}
A.~Dominguez, P.~Malgaretti, M.~N. Popescu, and S.~Dietrich.
\newblock Effective interaction between active colloids and fluid interfaces
  induced by marangoni flows.
\newblock {\em Phys. Rev. Lett.}, 116:078301, 2016.

\bibitem{blake1974}
J.~R. Blake and A.~T. Chwang.
\newblock Fundamental singularities of viscous flow. {P}art 1. the image system
  in the vicinity of a stationary no-slip boundary.
\newblock {\em J. Eng. Math.}, 8:23--29, 1974.

\bibitem{thutupalli2018}
S.~Thutupalli, D.~Geyer, R.~Singh, R.~Adhikari, and H.~A. Stone.
\newblock Flow-induced phase separation of active particles is controlled by
  boundary conditions.
\newblock {\em Proc. Nat. Ac. Sci. USA}, 2018.
\newblock (to appear).

\bibitem{kim2005}
Sangtae Kim and Seppo~J. Karrila.
\newblock {\em Microhydrodynamics}.
\newblock Butterworth-Heinemann, 1991.

\bibitem{blake1971}
J.~R. Blake.
\newblock A spherical envelope approach to ciliary propulsion.
\newblock {\em J. Fluid Mech.}, 46:199--208, 1971.

\bibitem{michelin2011}
S.~Michelin and E.~Lauga.
\newblock Optimal feeding is optimal swimming for all {P}\'eclet numbers.
\newblock {\em Phys. Fluids}, 23(10):101901, 2011.

\bibitem{soto2015}
R.~Soto and R.~Golestanian.
\newblock Self-assembly of active colloidal molecules with dynamic function.
\newblock {\em Phys. Rev. E}, 91:052304, 2015.

\bibitem{liebchen2018b}
B.~Liebchen and H.~L\"owen.
\newblock Which interactions dominate in active colloids?
\newblock 2018.
\newblock arxiv:1808.07389v1.

\bibitem{Jeffery1922}
G.~B. Jeffery.
\newblock The motion of ellipsoidal particles immersed in a viscous fluid.
\newblock {\em Proceedings of the Royal Society of London Series A},
  102(715):161--179, 1922.

\bibitem{pohl2014}
O.~Pohl and H.~Stark.
\newblock Dynamic clustering and chemotactic collapse of self-phoretic active
  particles.
\newblock {\em Phys. Rev. Lett.}, 112:238303, 2014.

\bibitem{jackson1962}
J.~D. Jackson.
\newblock {\em Classical Electrodynamics}.
\newblock John Wiley \& Sons, New York, 1962.

\bibitem{stone1996}
H.~A. Stone and A.~D.~T. Samuel.
\newblock Propulsion of microorganisms by surface distorsions.
\newblock {\em Phys. Rev. Lett.}, 77:4102, 1996.

\bibitem{liron1976}
N.~Liron and S.~Mochon.
\newblock Stokes flow for a stokeslet between two parallel flat plates.
\newblock {\em Journal of Engineering Mathematics}, 10(4):287--303, Oct 1976.

\bibitem{Spagnolie2012}
S.~E. Spagnolie and E.~Lauga.
\newblock Hydrodynamics of self-propulsion near a boundary: predictions and
  accuracy of far-field approximations.
\newblock {\em Journal of Fluid Mechanics}, 700:105--147, 2012.

\bibitem{Desreumaux2012}
N.~Desreumaux, N.~Florent, E.~Lauga, and D.~Bartolo.
\newblock Active and driven hydrodynamic crystals.
\newblock {\em The European Physical Journal E: Soft Matter and Biological
  Physics}, 35(8):1--11, 2012.

\bibitem{Tsang2014}
A.~C.~H. Tsang and E.~Kanso.
\newblock Flagella-induced transitions in the collective behavior of confined
  microswimmers.
\newblock {\em Physical Review E}, 90:021001, 2014.

\bibitem{Tsang2015}
A.~C.~H. Tsang and E.~Kanso.
\newblock Circularly confined microswimmers exhibit multiple global patterns.
\newblock {\em Physical Review E}, 91:043008, 2015.

\bibitem{Tsang2016}
Alan Cheng~Hou Tsang and Eva Kanso.
\newblock Density shock waves in confined microswimmers.
\newblock {\em Phys. Rev. Lett.}, 116:048101, Jan 2016.

\bibitem{Tsang2013}
A.~C.~H. Tsang and E.~Kanso.
\newblock Dipole interactions in doubly periodic domains.
\newblock {\em J. Nonlin. Sc.}, 23:971--991, 2013.

\bibitem{Coquereaux1990}
R.~Coquereaux, A.~Grossmann, and B.~E. Lautrup.
\newblock Iterative method for calculation of the weierstrass elliptic
  function.
\newblock {\em IMA Journal of Numerical Analysis}, 10:119--128, 1990.

\bibitem{Kanso2014}
E.~Kanso and A.C.H. Tsang.
\newblock Dipole models of self-propelled bodies.
\newblock {\em Fluid Dyn. Res.}, 46:061407, 2014.

\end{thebibliography}

\end{document}